\documentclass[11pt, a4paper]{article}
\usepackage{jheppub}
\usepackage[utf8]{inputenc}
\usepackage{lipsum, lmodern}
\usepackage{amsmath, amsthm, amssymb, amsfonts, braket, enumerate, slashed, upgreek, tensor, mathtools, enumitem}
\usepackage[most]{tcolorbox}
\usepackage[T1]{fontenc}
\usepackage{wrapfig, subcaption, graphicx, arydshln, float}
\usepackage{hyperref}
\usepackage{microtype}
\usepackage[font=small]{caption}
\graphicspath{ {./figures} }

\allowdisplaybreaks

\newcommand{\pa}{\partial}

\newcommand{\grav}{\text{grav}}
\newcommand{\eps}{\epsilon}

\newcommand{\dl}{\delta}
\newcommand{\nb}{\nabla}
\newcommand{\lie}{\mathsterling}
\newcommand{\Tr}{\operatorname{Tr}}

\newcommand{\A}{\mathcal{A}}
\newcommand{\mc}{\mathcal}

\newcommand{\dy}{d^{d-1}y\, }

\newcommand{\0}{{(0)}}
\newcommand{\1}{{(1)}}
\newcommand{\2}{{(2)}}

\newcommand{\RR}{\mathbb{R}}
\newcommand{\rbar}{\overline{r}}
\newcommand{\HH}{\mathbb{H}}
\newcommand{\matt}{\text{matter}}
\newcommand{\cw}{\curlywedge}
\newcommand{\ba}{\mathsf{a}}
\newcommand{\bF}{\mathsf{f}}

\newcommand{\bbF}{\mathsf{F}}
\newcommand{\hpsi}{\widehat{\psi}}
\DeclareMathOperator{\Disc}{\operatorname{Disc}}

\title{Modular Witten diagrams and quantum extremality}
\author[a]{Abhirup Bhattacharya,}
\author[a]{Onkar Parrikar}
\affiliation[a]{Department of Theoretical Physics, Tata Institute of Fundamental Research,\\
                Mumbai 400005, India}
\emailAdd{abhirup.bhattacharya@tifr.res.in}
\emailAdd{parrikar@theory.tifr.res.in}

\abstract{We study entanglement entropy for ball-shaped regions in excited states of holographic conformal field theories. The excited states are prepared by the Euclidean path integral in the CFT with a source turned on for some double-trace operator, with a small, $O(1)$ amplitude $\lambda$. On the gravity side, the double-trace operator deforms the bulk geometry as well as the entanglement structure of the state of bulk matter fields. By the quantum extremal surface formula, this leads to a deformation of the shape of the entanglement wedge, an effect which becomes manifest in the entanglement entropy at $O(\lambda^2 G_N)$. On the CFT side, we explicitly calculate the entanglement entropy perturbatively in the source amplitude to $O(\lambda^2)$, in terms of modular-flowed correlation functions of double-trace operators. We then evaluate these modular-flowed correlation functions using Witten diagrams. This calculation involves a Schwinger-Keldysh contour ordering prescription in the bulk, which we motivate using analytic continuation from Euclidean replica correlators. Focusing on a particular graviton-exchange diagram, we rewrite it in a form where it manifestly reproduces the canonical energy term present in the quantum Ryu-Takayanagi formula, including the shape deformation of the entanglement wedge due to backreaction and quantum effects.  }

\begin{document}

\preprint{\parbox{3cm}{TIFR/TH/25-23}}
\maketitle

\parskip=10pt

\section{Introduction}
The quantum Ryu-Takayanagi formula for entanglement entropy in a holographic conformal field theory (dual to Einstein gravity coupled to matter) states that the entanglement entropy of a boundary subregion $R$ in a holographic CFT state $\Psi$ is given by \cite{Ryu:2006bv, Hubeny:2007xt, Faulkner:2013ana, Engelhardt:2014gca}
\begin{equation} \label{eq:QEE}
S_{R}^{\Psi}= \text{ext}_X\,S^{g,\psi}_{\text{gen}}(X),\qquad S^{g,\psi}_{\text{gen}}(X)= \frac{1}{4G_N}\A^{g}(X) + S^{g,\psi}_{\text{bulk}}(X),
\end{equation}
where $g$ is the bulk spacetime metric, $\psi$ is the state of bulk matter fields, $\mathcal{A}$ is the area of a co-dimension two surface $X$, and the extremization is over all surfaces $X$ which are anchored on $\pa R$ and homologous to $R$. The RT formula has been tremendously important with many important applications. However, it is perhaps fair to say that from a Lorentzian or Hilbert space point of view, the origins of this formula are rather poorly understood. The seminal derivation of Lewkowycz and Maldacena \cite{Lewkowycz:2013nqa}, for instance, uses the Euclidean path integral of quantum gravity, which obscures the Hilbert space interpretation. In very special symmetric situations such as ball-shaped regions in the global vacuum state, one can appeal to locality of modular flow \cite{Casini:2011kv}, but this case is highly non-generic. Tools from the theory of quantum error correction have been used to argue that something like an RT formula is a natural consequence of subregion duality in toy models \cite{Almheiri:2014lwa, Dong:2016eik, Harlow:2016vwg}, but those insights have not directly translated to a Hilbert space derivation of the RT formula. Recent work on algebras in the large $N$ limit has shown why the generalized entropy is natural from the point of view of gravitationally dressed observables \cite{Witten:2021unn, Chandrasekaran:2022cip, Chandrasekaran:2022eqq, Jensen:2023yxy}, but not explained the extremization over surfaces in equation \eqref{eq:QEE}. Having said all this, it is perhaps too much to expect a Hilbert space derivation of equation \eqref{eq:QEE} directly in gravity -- after all, the area term comes from the $O(N^2)$ entanglement between the underlying microscopic degrees of freedom. Nevertheless, one might hope to better understand quantum gravity if we could find some explanation for why gravity knows about the entanglement between microscopic degrees of freedom in terms of such a simple and elegant, geometric formula. 

One approach has been to look at the behavior of entanglement under macroscopic changes in the boundary state. For instance, it was shown in \cite{Faulkner:2013ica} that for a classical change $\delta \Psi$ is the boundary CFT state (i.e., corresponding to $O(N^2)$ deformation of the sources in the path integral), the linearized change in the entanglement entropy for ball-shaped regions is given by the first law of entanglement, which upon using the bulk Einstein equations precisely matches with the RT formula. This was then generalized to quantum deformations of the state in \cite{Swingle:2014uza}. At this order, however, one does not see the deformation of the surface due to the extremality condition; that appears at second order in the CFT state perturbation. In \cite{Faulkner:2017tkh, Haehl:2017sot}, it was shown that for classical sources, the second order perturbation of the boundary entanglement entropy precisely matches with the corresponding bulk formula, including the change in the extremal surface. Interestingly, the change in the surface stems from some novel boundary terms which arise from integrated modular flowed correlation functions of the stress tensor. The purpose of the present paper is to generalize the calculations of \cite{Faulkner:2017tkh} to the case of sources which deform the bulk geometry \emph{and} matter entanglement structure, and show the emergence of the quantum extremality formula from modular-flowed correlation functions which arise in the second order perturbation of entanglement entropy in the CFT (see also \cite{Haehl:2019fjz, Belin:2021htw, Chowdhury:2024fpd} for previous work along these lines). Specifically, we will turn on $O(1)$ sources for double-trace operators in the CFT and study the corresponding change in the entanglement entropy to second order, which involves, among other things, modular-flowed correlators of double-trace operators. In computing such correlation functions in holography, we are naturally led to consider Witten diagrams for modular-flowed correlators, or \emph{modular Witten diagrams}, which are very similar in spirit to the Witten diagrams encountered in the gravity dual of Schwinger-Keldysh correlators \cite{Glorioso:2018mmw, Jana:2020vyx}. In this paper we only need to study modular Witten diagrams for the case of local modular flow, but it might also be of independent interest to study modular Witten diagrams for the case of non-local modular flow (see section \ref{sec:grsk-derivation} for further discussion). 

While the point of the present paper is to confirm aspects of the quantum RT formula from a Lorentzian calculation in AdS/CFT (see also \cite{Lewkowycz:2018sgn, Soni:2024oim} for previous work in this direction), an underlying motivation is also to develop tools to further explore quantum gravity corrections to this formula. For instance, very little is known about the role of gravitons in the quantum RT formula (although see \cite{Colin-Ellerin:2025dgq} for some recent progress), and accommodating these corrections could lead to important new additions to the quantum RT formula.\footnote{We thank Shiraz Minwalla for many illuminating conversations on this subject.} We believe that the tools developed in this paper can be used to explore this question, for instance, by including graviton loops in modular Witten diagrams. However, this will not be attempted in the current work.  

\subsection{Setup and outline}
Throughout this paper we will work with conventions where the boundary CFT lives in $d$ spacetime dimensions and the bulk gravity theory lives in $d+1$ spacetime dimensions. On the CFT side we consider a one-parameter family of Euclidean path integral states $\Psi(\lambda)$ generated by turning on an $O(1)$ source for a double-trace operator in the Euclidean path integral. 
At $\lambda = 0$, $\Psi$ reduces to the vacuum state, and the deformation induces an $O(1)$ stress tensor in the bulk which back-reacts on the geometry. Accordingly, on the bulk side, we consider a one-parameter family of asymptotically AdS geometries $g_{ab}(\lambda)$ with the corresponding matter fields being in a one-parameter family of states $\psi(\lambda)$. The pair $(g_{ab}(\lambda), \psi(\lambda))$ satisfies the semi-classical Einstein equation
\begin{equation} \label{eq:SCEE}
R_{ab}(\lambda) - \frac{1}{2}R(\lambda)\,g_{ab}(\lambda) + \Lambda\, g_{ab}(\lambda) = \langle T_{ab}\rangle_{g,\psi}(\lambda).
\end{equation}
Throughout this work, we will ignore graviton fluctuations and only consider back-reaction up to $O(G_N)$. This semiclassical approximation may be interpreted as the gravitons being in a coherent state, with the matter fields occupying a non-coherent state. 

Now, consider a subregion $R$ in the boundary CFT. Then the entanglement entropy associated to $R$ in the state $\Psi$ is given by\footnote{In this paper we will assume that the QFT Hilbert space is factorizable. The correct way to think about the problem is in terms of von Neumann algebras of operators associated to spacetime subregions. For QFT these algebras are of type-III \cite{Witten:2018zxz}, which means that one cannot assign reduced density operators and traces to subregions. We will not worry about these subtleties in this work, but all of our CFT calculations can be reworded in the more precise language of Tomita-Takesaki theory for type-III algebras.}
\begin{equation}
    \rho_R^{\Psi}(\lambda) = \Tr_{\bar{R}} | \Psi(\lambda) \rangle \langle \Psi(\lambda) |, \qquad S_R^{\Psi}(\lambda) = - \Tr_R \rho_R^{\Psi}(\lambda) \log \rho_{R}^{\Psi}(\lambda).
\end{equation}
The quantum Ryu-Takayanagi formula \cite{Engelhardt:2014gca} states that the entanglement entropy of $R$ is given by the generalized entropy across the quantum extremal surface in the bulk. In order to find this surface, we are instructed to extremize the \emph{generalized entropy} functional:
\begin{equation} 
S_{R}^{\Psi}= \text{ext}_X\,S^{g,\psi}_{\text{gen}}(X),\qquad S^{g,\psi}_{\text{gen}}(X)= \frac{1}{4G_N}\A^{g}(X) + S^{g,\psi}_{\text{bulk}}(X),
\end{equation}
with respect to the surface $X$, where $\A$ is the area of $X$ and $S_{\text{bulk}}$ is the entropy of bulk quantum fields across $X$. Note that for every value of $\lambda$, we get a distinct quantum extremal surface. This defines a one-parameter family of surfaces, which in some convenient set of coordinates can be described by the embedding functions $x^a(y;\lambda)$ where $y^\alpha$ are a set of intrinsic coordinates on the surface, and $x^a$ are the ambient coordinates. We will refer to the extremal surface at $\lambda = 0$ as the \emph{undeformed QES} $X^\0$, and the extremal surface at finite values of $\lambda$ as the \emph{deformed QES}.

\begin{figure}[t]
   \centering
   \includegraphics[height=6cm]{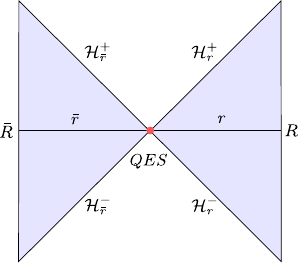}
   \caption{A cross section of the left and right Rindler wedges dual to $\bar{R}$ and $R$ respectively. The shaded regions are the domains of dependence $D(r)$ and $D(\bar{r})$ of the homology surfaces $r$ and $\bar{r}$ respectively. The red dot at the center is the undeformed QES sitting at $x^\pm = 0$ and $\mc{H}^{\pm}$ are the null horizons.}
   \label{fig:rindler}
\end{figure}
In this work, we will restrict ourselves to cases where the subregion $R$ in the background state $\Psi^\0 \equiv \Psi(\lambda = 0)$ has local modular flow. For concreteness, we will take $R$ to be either a half-space (or equivalently a ball-shaped region \cite{Casini:2011kv}), and $\Psi^{(0)}$ to be the CFT vacuum:
\begin{equation}
    R = \left\{ x \in \RR^d\, |\, x^0 = 0, x^1 \geq 0  \right\}.
\end{equation}
The corresponding bulk geometry $g^\0$ is pure AdS, and the entanglement wedge of $R$ in the state $\Psi^\0$ is the AdS Rindler wedge which can be parametrized in the following way \cite{Casini:2011kv}:
\begin{equation} \label{eq:Rindler}
    g^\0 = -(\rho^2 - 1)dt^2 + \frac{d\rho^2}{\rho^2 - 1} + r^2 d\Omega_{d-1}^2.
\end{equation}
The undeformed QES lies on the surface $\rho = 0$ and $t = 0$. It is convenient to work with surface adapted light cone coordinates given by $x^\pm = \sqrt{\rho^2 - 1} e^{\pm t}$. In these new coordinates the undeformed QES is given by $X^\0 = \{ x \in M\, |\, x^\pm = 0  \}$ (see appendix \ref{app:A} for more details). We will often use $r$ to denote a homology surface between $R$ and $X$, the corresponding entanglement wedge is its domain of dependence $D(r)$. For concreteness, we will choose $r$ and $\bar{r}$ to be time reflection symmetric slices in the coordinates of eq.\ \eqref{eq:Rindler}. At $\lambda = 0$, the bulk state of quantum fields $\psi^\0$ is the vacuum, and the bulk modular flow associated to $D(r)$ is also local and is generated by the vector field
\begin{equation} \label{eq:xi}
    \xi = 2\pi(x^{+} \partial_{+} - x^{-} \partial_{-}). 
\end{equation}
It is easy to check that $\xi$ asymptotically approaches the boost generator on the boundary. 

The goal of the present paper is to calculate the entanglement entropy of the double-trace deformed boundary state to $O(\lambda^2)$, and show that it reproduces the quantum extremal corrections to the shape of the entanglement wedge. In particular, our main motivation here is to see how the \emph{quantum} extremality formula emerges from the boundary calculation, thus naturally generalizing the calculations in \cite{Faulkner:2017tkh}. Concretely, we compute the relative entropy between the excited state $\Psi(\lambda)$ and the vacuum, which is related to the change in the entanglement entropy via the standard formula \( S_{\text{rel}}(\Psi \| \Omega) = \dl \langle K \rangle - \dl S_{\text{EE}}\). Phrasing the discussion in terms of relative entropy has a couple of advantages. Unlike the entanglement entropy, the relative entropy is UV finite. Moreover, as we explain in section \ref{sec:gravity}, it is naturally dual to (a semiclassical generalization of) the gravitational symplectic form plus an additional boundary term which encodes all information about the displacement profile of the QES \cite{Lashkari:2015hha, Faulkner:2017tkh}. 

\begin{figure}[t]
    \centering
    \includegraphics[width=0.33\linewidth]{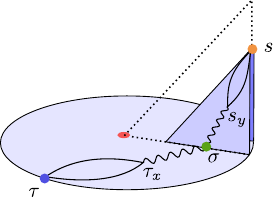}
    \caption{Inserting a complete set of states in the $s$-channel diagram with one vertex in the Euclidean segment and the other vertex in the Lorentzian segment. The green dot denotes the location on the $t=0$ slice where the bulk-to-bulk graviton propagator is cut.}
    \label{fig:s-main}
\end{figure}

The rest of the paper is organized as follows: in section \ref{sec:QES}, we will first obtain a concrete formula for the deformation of the extremal surface in the bulk due to quantum extremal effects. In section \ref{sec:gravity}, we will use this to calculate the $O(\lambda^2)$ correction to the relative entropy using the quantum RT formula. In section \ref{sec:SK}, we then turn to the CFT calculation of the relative entropy and write it in terms of a certain modular-flowed correlation function of the double trace operator that deforms the CFT state. We will then use Witten diagrams to evaluate this modular-flowed correlation function and rewrite it in a way such that the quantum extremal corrections to the entanglement wedge become manifest. In the following, we will primarily focus on a particular graviton exchange diagram (see figure \ref{fig:s-main}) which is responsible for reproducing the local geometric terms in our main gravitational formula eq.\ \ref{eq:SrelEq}. This will be enough for us to recover the QES prescription. As we emphasize in section \ref{sec:SK}, there are other Witten diagrams corresponding to multiparticle exchanges which we do not process in this paper. Calculating these diagrams will be important to systematically evaluate all $O(G_N)$ corrections to the boundary relative entropy, including the relative entropy of bulk matter. We leave this for future work.

\section{Displacement profile of the quantum extremal surface}\label{sec:QES}
Our main goal in this section is to derive an equation for the displacement profile (i.e., the shape) of the quantum extremal surface (QES) under a state deformation in the CFT, following \cite{Levine:2020upy}. The novelty in this calculation is that one must include quantum extremal corrections coming from the change in the state of bulk matter fields. We will give a general formula for the displacement profile for a half space cut (dual to the AdS Rindler wedge in the bulk) which is valid up to $O(G_N)$. 

\subsection{Displacement profile at $O(G_N)$} \label{sec:extOall}
We consider the boundary half-space cut in the state $\Psi(\lambda)$. As discussed before, at any value of $\lambda$, the state of bulk matter fields $\psi(\lambda)$ gives rise to an $O(1)$ stress tensor which will back-react on the metric. However, since we are interested in perturbation theory in $G_N$, we can regard the change in the metric $\delta g_{ab}(\lambda)$ as being small -- in particular of $O(G_N)$. In this section, we will derive a formula for the change in the shape of the extremal surface that is valid up to $O(G_N^2)$ errors and \emph{all orders} in $\lambda$. Eventually in section \ref{sec:extOlambda} we will perform a double expansion in both $\lambda$ and $G_N$ and study infinitesimal shape deformations at $O(\lambda G_N)$.\footnote{Further details on this double expansion can be found in section \ref{sec:REGrav} -- see the discussion around eq.\ \eqref{eq:SCexp}.} In section \ref{sec:SK} we will rederive the $O(\lambda G_N)$ formula from a CFT calculation. 

At any finite value of $\lambda$, this change in the metric is obtained by solving the linearized Einstein equation around the background $g^{(0)}$: 
\begin{equation} \label{eq:SCLEE}
E^{(1)}_{ab}(\delta g) = 8\pi G_N \Delta\langle T_{ab} \rangle_{g^{(0)},\psi},
\end{equation}
where $E^{(1)}$ is the linearized Einstein tensor (including the cosmological constant term), and $\Delta\langle T_{ab} \rangle$ is the vacuum subtracted bulk stress tensor at a \emph{finite} value of $\lambda$.\footnote{Throughout this section, we will use $\Delta$ to denote a finite difference between $\Psi(\lambda)$ and $\Psi^\0$, while $\dl$ will be reserved for $O(G_N)$ differences.} Under the state deformation, we expect the undeformed extremal surface to move to the location $X(\lambda) = \{x\in M\ |\ x^\pm =  v^{\pm}(y) + O(G_N^2)\}$, where $v^\pm$ is the deformation of the QES to $O(G_N)$; we will call this the \emph{displacement profile}. The location of this new surface is determined by the QES prescription:
\begin{equation} \label{eq:QES_def}
\dl_X \left( \frac{1}{4G_N}\A^g + S^{g,\psi}_{\text{bulk}}\right)_{X=X(\lambda)} = 0.
\end{equation}
The strategy now is to expand the quantity inside brackets to $O(G_N)$ and derive the extremality equation satisfied by $v^{\pm}$. In this subsection, we will solve the extremality equation non-perturbatively in $\lambda$, and in subsection \ref{sec:extOlambda} we will specialize to the case where $v$ is also $O(\lambda)$. 


To begin with, we expand eq. \eqref{eq:QES_def} by writing $X= X^{(0)} + v + \delta X$, where $\delta X$ is the variation and $v$ is the displacement profile. We need only keep terms to $O(v\, \delta X)$:
\begin{equation} \label{eq:QESVariation0}
    \dl_X \left[\frac{1}{4G_N} (\dl_v + \dl_g) \A^g + (\dl_v + \dl_g + \Delta_\psi) S^{g,\psi}_{\text{bulk}} \right]_{X=X^\0} = 0,
\end{equation}
where $\Delta_{\psi}$ denotes the difference between the vacuum state $\psi^\0$ and $\psi(\lambda)$. Let us first evaluate the area terms in eq. \eqref{eq:QESVariation0}. The area functional for a general shape deformation $\delta X^I$ is given by
\begin{align}
    \begin{split}
        \A(X + \dl X) &= \int \dy \sqrt{\det(g_{\alpha\beta} + g_{IJ}\partial_{\alpha} \dl X^I                       \partial_{\beta} \dl X^J)}\\
                         &= \A(X^\0) + \frac{1}{2} \int \dy \sqrt{\gamma} \gamma^{\alpha\beta} g_{IJ}\partial_{\alpha}\dl X^I \partial_{\beta} \dl X^J+\cdots,
    \end{split}
\end{align}
where $\gamma_{\alpha\beta}$ is the induced metric on the undeformed extremal surface $X^\0$. Note that since the surface $X^\0$ is already classically extremal, there is no $O(\dl X^I)$ term. We can specialize this to the case where $\delta X^I = v^I + \delta x^I$, where the first term is the $O(G_N)$ displacement coming from back-reaction due to the state deformation, and $\delta x^I$ is the shape variation $\delta_X$ in equation \eqref{eq:QESVariation0} coming from the extremality condition. Integrating by parts and discarding boundary terms at infinity\footnote{This is allowed because $v^I$ vanishes at the asymptotic boundary, as the surface is still anchored at $\pa A$.}, we arrive at
\begin{equation} \label{diffOP}
    \dl_v \dl_X \A^g(X^\0) = -\int \dy \dl x^I \partial_{\alpha}(\sqrt{\gamma}\, \gamma^{\alpha\beta}g_{IJ} \partial_{\beta} v^J).
\end{equation}
Next, let us look at the change in $\A$ under the metric perturbation. Taking a metric variation of the area functional around a general initial surface $X$, we obtain
\begin{equation}
    \dl_g \A^g(X) = \frac{1}{2} \int \dy \sqrt{\gamma}\, \gamma^{\alpha\beta} \partial_{\alpha} X^a \partial_{\beta} X^b \dl g_{ab}.
\end{equation}
Taking a further shape derivative and evaluating on the original QES yields
\begin{align} \label{eq:dlXdlgARindler}
    \begin{split}
        \dl_X\dl_g \A^g(X^\0) &= \frac{1}{2}\int \dy \sqrt{\gamma}\, \gamma^{\alpha\beta}\left( 2 \dl g_{I\beta}                           \partial_{\alpha} \dl x^I + \dl x^I \partial_I \dl g_{\alpha\beta} \right)\\
                         &= \int \dy \dl x^I \left[ -\partial_{\alpha}(\sqrt{\gamma}\, \gamma^{\alpha\beta} \dl g_{I\beta}) + \frac{\sqrt{\gamma}}{2} \gamma^{\alpha\beta} \partial_I\dl g_{\alpha\beta}  \right].
    \end{split}
\end{align}
Now we come to the bulk entropy terms. To process $\dl_X S_{\text{bulk}}$ we use the trick of rewriting the entanglement entropy in terms of the relative entropy and the expectation value of the modular Hamiltonian. Let $K^\0_r$ be the modular Hamiltonian associated to the undeformed entanglement wedge for the state $\psi^\0$. Then we have
\begin{equation} \label{eq:Srel}
    \dl_X S^{g,\psi}_{\text{bulk}}(X^\0) = - \dl_X S_{\text{rel}}(\psi \| \psi^\0) + \dl_X \braket{K^\0_r}_{\psi}.
\end{equation}
A general formula for the shape derivative of relative entropy was derived by Ceyhan and Faulkner in \cite{Ceyhan:2018zfg} in the context of deriving the quantum null energy condition using the methods of Tomita-Takesaki theory. This is is the so-called \emph{ant formula} which goes as follows:
\begin{equation} \label{eq:CF}
    \frac{\dl S_{\text{rel}}(\psi \| \psi^\0)}{\dl X^{\pm}(y)} = -\inf_{s} \braket{\psi_s | P_\pm(y) | \psi_s},
\end{equation}
where $P_\pm(y)$ are the averaged null energy (ANE) operators defined as
\begin{equation}
    P_+(y) = 2\pi \int_{-\infty}^{\infty}dx^+\, T_{++}(x^+, 0, y),
\end{equation}
and $\ket{\psi_s}$ is the Connes cocycle flowed state $\ket{\psi_s} = u^\prime_{\psi^\0|\psi}(s) \ket{\psi}$ in the bulk\footnote{A proof can be found in the original reference \cite{Ceyhan:2018zfg}. A simpler proof under slightly more restrictive assumptions can be found in \cite{Hollands:2025glm}. For more details on the Connes cocycle see appendix \ref{app:CC}}. Note that the prime on the Connes cocycle refers to the fact that the flow is taken over the complementary Rindler wedge $D(\bar{r})$. Plugging eq. \eqref{eq:CF} into eq. \eqref{eq:Srel} and acting on both sides with $\Delta \equiv (\Delta_\psi + \dl_g)$, we get
\begin{equation}
    \dl_X \Delta S^{g,\psi}_{\text{bulk}}(X^\0) = \inf_{s} \Delta \braket{P_+(y)}_{\psi_s} + \Delta \braket{\dl_X K^\0_r}_{\psi},
\end{equation}
where, for simplicity, we have considered the shape derivative $\dl_X$ is along the $x^+$ direction, but similar formulas can also be written for the $x^-$ direction. Now, it was shown in \cite{Faulkner:2016mzt} that the shape derivative of the modular Hamiltonian is proportional to the half-sided ANE operators. Concretely, we have\footnote{Note that this expression is valid only if we assume factorization of the underlying Hilbert space. Without factorization one should replace the left hand side of this equation with the logarithm of the full modular operator $-\log \Delta_\psi$ and the right hand side with the full ANE operator. In factorized setting this is precisely the difference between the left and right modular Hamiltonians. Nevertheless, in this paper we will assume factorization and work with the one-sided formula.}
\begin{equation} \label{eq:MHSD}
    \frac{\dl}{\dl X^+(y)} \braket{K^\0}_{\psi} = -\braket{P_{+;r}(y)}_{\psi},
\end{equation}
where $P_{+;r}(y)$ is the half-sided ANE operator supported only on the future null horizon $\mc{H}^{+}_{r}$ of $D(r)$ (see figure \ref{fig:rindler}):
\begin{equation}
    P_{+;r}(y) = 2\pi \int_{0}^\infty dx^+\, T_{++}(x^+, 0, y).
\end{equation}
Eq. \eqref{eq:Srel} then becomes
\begin{equation}
    \dl_X \Delta S^{g,\psi}_{\text{bulk}}(X^\0) = \inf_{s} \Delta \braket{P_+(y)}_{\psi_s} -  \Delta\braket{P_{+;r}(y)}_\psi.
\end{equation}
Finally, note that the $\delta_v\delta_xS_{\text{bulk}}$ term contributes to the homogenous part of the equation for $v^I$ and modifies the differential operator (see equation \eqref{diffOP}) acting on $v^I$ into an integral operator. However, since it contributes to the homogenous part, and further since its contribution is down by $G_N$ as compared to the contribution coming from equation \eqref{diffOP}, we can ignore this term at leading order in $G_N$.

Collecting everything together we arrive at the following extremality equation for the displacement profile for the QES:
\begin{align} \label{eq:QEEall}
    \begin{split}
        \widehat{\nabla}^2 v_+ &= -\left(\widehat{\nabla}_{\alpha} \dl g\indices{^\alpha_+} - \frac{1}{2}\nabla_+ \dl g\indices{^\alpha_\alpha}\right) + 8\pi G_N \inf_{s} \int_{-\infty}^\infty dx^+ \Delta \langle \psi_s | T_{++}(x^+, 0, y) | \psi_s\rangle\\ &- 8\pi G_N \int_0^\infty dx^+ \Delta \langle \psi | T_{++}(x^+, 0, y) | \psi\rangle + \cdots,
    \end{split}  
\end{align}
where the $\cdots$ denote terms that are $O(G_N^2)$. A similar equation holds for $v_-$. We can process this equation a bit further by using the fact that the metric perturbation satisfies the linearized Einstein equation which, when evaluated on the null horizon defined by $x^- = 0$ reads
\begin{equation} \label{eq:linEEOp}
    -\frac{1}{2} \nabla^2 \hat{h}_{++} - \frac{1}{2} \nabla_+ \nabla_+ \hat{h}\indices{^a_a} + \nabla_a \nabla_+ \hat{h}\indices{^a_+} + d \hat{h}_{++} = 8\pi G_N T_{++}.
\end{equation}
Note that we are thinking of the above as an operator equation satisfied by the linearized graviton operator $\hat{h}_{ab}$. Taking the expectation value of the above linearized equation of motion in the state $\psi$ (or equivalently $|\psi_s\rangle = u^\prime_{\psi|\psi^\0}(s) | \psi\rangle$) we find that we can identify the metric perturbation $\dl g_{ab}$ in the semiclassical Einstein equation eq. \eqref{eq:SCLEE} with the quantity $\Delta \langle \psi | \hat{h}_{ab} | \psi\rangle$. With this identification, we now plug in eq. \eqref{eq:linEEOp} into eq. \eqref{eq:QEEall}. Performing the $x^+$ integrals and isolating the boundary terms localized at $X^\0$ we find
\begin{equation} \label{eq:QEEall2}
    \widehat{\nabla}^2 v_+ = -\frac{1}{2} \widehat{\nabla}^2 \inf_{s} \int_{-\infty}^{\infty}dx^+\, \Delta \langle\psi_s | \hat{h}_{++}(x^+, 0, y) | \psi_s\rangle + \frac{1}{2} \widehat{\nabla}^2 \int_0^\infty dx^+\, \Delta \langle\psi | h_{++}(x^+, 0, y) | \psi \rangle.
\end{equation}
Note that the boundary terms localized on the QES have exactly canceled the $\dl_X \dl_g \mc{A}$ term in eq. \eqref{eq:QEEall}. Since the Connes cocycle acts as a one-sided boost in the complementary Rindler wedge $D(\bar{r})$ at the level of one-point functions, we see that the expectation value of the graviton in the flowed state is the same as the expectation value in the un-flowed state in the region $D(r)$, \emph{i.e.}
\begin{equation}
    \Delta \langle\psi_s | h_{++}(x^+, 0, y) | \psi_s\rangle = \Delta \langle \psi | h_{++}(x^+, 0, y) | \psi\rangle \quad \text{when} \quad x^+ > 0.
\end{equation}
Therefore the terms in eq. \eqref{eq:QEEall2} supported on the future null horizon $\mc{H}^{+}_r$ drop out. Finally, the extremality equation becomes
\begin{equation}
    \widehat{\nabla}^2 v_+ = -\frac{1}{2} \widehat{\nabla}^2 \inf_{s} \int_{-\infty}^0 dx^+\, \Delta \langle \psi_s | \hat{h}_{++}(x^+, 0, y) | \psi_s\rangle.
\end{equation}
Note that $v_+ \to 0$ asymptotically because the surface is always anchored at the same place. In addition, $\Delta \langle h_{++} \rangle_{\psi_s}$ also goes to zero asymptotically, as there are no sources for the CFT stress tensor turned on in the Lorentzian section. This lets us solve for $v$:
\begin{equation} \label{eq:vplusAll}
    v_+(y) = -\frac{1}{2} \inf_s \int_{-\infty}^0 dx^+\Delta \langle\psi_s | \hat{h}_{++}(x^+, 0, y) | \psi_s\rangle.
\end{equation}
A similar equation holds for $v_-$. This equation is valid up to $O(G_N^2)$ corrections and at all orders in $\lambda$.

\begin{figure}[t]
    \centering
    \includegraphics[width=\linewidth]{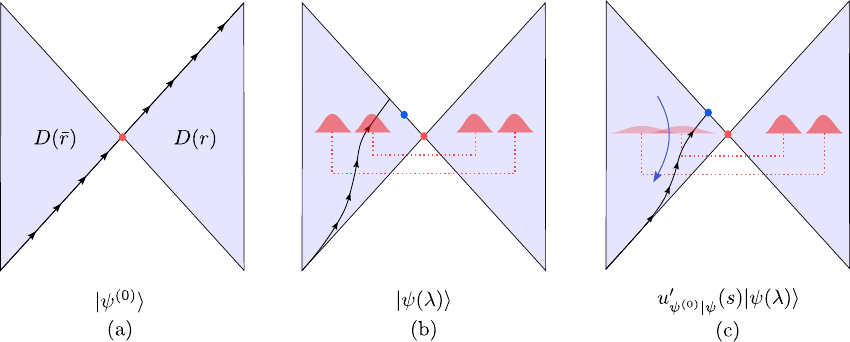}
    \caption{(a) The undeformed Rindler wedge in the background state $\psi^\0$. The black line with arrows denotes the generator along the null horizons $\mc{H}_r^+ \cup \mc{H}_{\bar{r}}^{-}$. The red dot at the center is the undeformed QES. (b) The red bumps denote matter excitations in the state $\psi(\lambda)$ which may be entangled between the two wedges, as indicated by the red dotted lines. The arrowed black line denotes the deformed null generators and the blue dot is the deformed QES. (c) The bulk matter state under CC flow in the left wedge $\bar{r}$. The CC flow dilutes away all matter excitations in $D(\bar{r})$, and time delay of the null generator is minimized in the $s\rightarrow \infty$ limit.}
    \label{fig:all-orders}
\end{figure}

This equation has a nice geometrical interpretation: at $O(G_N)$, one can think of $\delta g_{ab} = \Delta\langle \psi| \hat{h}_{ab}|\psi\rangle $ as the back-reaction on the metric due to the matter stress tensor. Correspondingly, the quantity 
\begin{equation}
    v_+^{\text{causal}} = -\frac{1}{2} \int_{-\infty}^0 dx^+\Delta \langle\psi | \hat{h}_{++}(x^+, 0, y) | \psi \rangle
\end{equation}
is the time-delay that a null ray traversing along $\mathcal{H}^{-}_{\bar{r}}$ suffers due to the matter sources. As a result, this null ray will not in general end up on the true displaced QES; the true QES generally lies behind the causal horizon. However, by acting on the state $\psi$ with the Connes cocyle flow in $\bar{r}$, we can move things around in $\bar{r}$, while leaving the QES unaffected. In equation \eqref{eq:vplusAll}, by going to the value of $s$ where the infimum is reached, one is trying to minimize the time delay suffered by the ingoing null ray in $D(\overline{r})$, thus trying to push it as ``inwards'' as possible. The content of equation \eqref{eq:vplusAll} is then that by doing so, one arrives at the true QES. As alluded to in figure \ref{fig:all-orders}, the exact mechanism by which the time delay is minimized is dilution of the matter excitations. In the strict $s \to \infty$ limit all fluctuations in the matter stress tensor are completely diluted out, and any local observer on the left wedge perceives a state that is indistinguishable from the vacuum. A numerical demonstration of this in the context of AdS$_3$ is given in appendix \ref{app:CCBTZ}.

\subsection{Displacement profile at $O(\lambda)$} \label{sec:extOlambda}
We will now specialize to the case of infinitesimal state perturbations. The resulting formula for the displacement profile will be valid to $O(G_N)$ and to $O(\lambda)$ in the state deformation. Going back to equation \eqref{eq:QEEall2}, we see that the first term of the right hand side vanishes at linearized order in $\lambda$: this is because at linearized order in state perturbations, we have the operator $\inf \int_{-\infty}^{\infty}dx^+ \widehat{\nabla}^2h_{++}(x^+,0,y)$ acting on the background state either on the ket side or on the bra side. Since the ANEC operator annihilates the background state, linearized Einstein equation implies that the above operator must also annihilate it. So, we are left with:
\begin{equation} \label{eq:QEE_Rindler}
    \widehat{\nabla}^2 v_+ = \widehat{\nabla}^2 w_+,
\end{equation}
where we have defined
\begin{equation}
    w_+(y) = \frac{1}{2}\int_{0}^{\infty} dx^+ \delta g_{++}(x^+, 0, y).
\end{equation}
Since both $v_+$ and $w_+$ asymptotically decay to zero at spatial infinity, we can solve for $v_+$:
\begin{equation} \label{eq:vplus}
    v_+(y) = w_+(y) = \frac{1}{2} \int_0^\infty dx^+\, \dl g_{++}(x^+, 0, y).
\end{equation}
A similar formula holds for $v_-$. So we have arrived at a formula for the shape deformation of the QES at linear order in the state deformation, up to $O(G_N^2)$ corrections. Note that this formula incorporates quantum extremal effects, i.e., the effect of the bulk entropy term in the generalized entropy on the shape of the extremal surface. Our goal throughout the rest of the paper will be to reproduce eq. \eqref{eq:vplus} from a first principles CFT computation, together with the standard Witten diagrammatic rules of AdS/CFT. This will serve as a verification of the quantum extremality prescription at $O(G_N)$ and $O(\lambda)$. But first, we will see how the displacement profile enters the entanglement entropy calculation using the quantum RT formula.

\section{Bulk relative entropy using the quantum RT formula}\label{sec:gravity}
We will now derive an explicit bulk formula for the relative entropy between the excited state and the vacuum state at $O(\lambda^2)$ using the QES prescription. In doing so, we will make extensive use of the gravitational covariant phase space formalism, suitably generalized to semi-classical gravity \cite{Bhattacharya:2025cld}. We will first begin by reviewing some aspects of this formalism.

\subsection{Review of the covariant phase space formalism}
In this subsection, we will provide a brief review of the covariant phase space formalism, first in the classical case as developed in \cite{crnkovic1987covariant, Lee:1990nz, Wald:1993nt, Iyer:1994ys, Iyer:1995kg, Wald:1999wa}, and then in semiclassical gravity \cite{Bhattacharya:2025cld}. We will focus on the case of Einstein gravity coupled to matter in asymptotically AdS spacetimes.

\subsubsection*{Classical case}
We define the \emph{configuration space} $\mc{C}$ of the theory as the space of all possible fields on the spacetime manifold $M$ consistent with appropriate asymptotic boundary conditions at infinity. The standard choice of boundary conditions in gravity are Dirichlet boundary conditions, where the values of the metric and matter fields are fixed on $\partial M$. With this choice of boundary conditions, the Lagrangian of our theory is given by
\begin{equation}
    L[g, \phi^A] = L_{EH}[g] + L_{\matt}[g, \phi^A] + d \ell,
\end{equation}
where $L_{EH}$ is the Einstein-Hilbert Lagrangian, $L_{\matt}$ is the matter Lagrangian, and $\ell$ is an arbitrary smooth extension of the Gibbons-Hawking-York boundary term to the bulk, along with counterterms necessary for holographic renormalization. The specific choice of this extension will not be important. Note that it is conventional in this formalism to think of the Lagrangian as a top-degree form on $M$.

We will assume that $\mc{C}$ is endowed with sufficient structure so that standard techniques of differential calculus apply \cite{crnkovic1987covariant, Harlow:2019yfa}. In particular, we will assume the existence of an exterior derivative $\dl$ on $\mc{C}$, which can be thought of as an anti-symmetrized variation of all (dynamical) fields in $L$ satisfying $\dl^2 = 0$. Varying $L$ gives
\begin{equation} \label{eq:dlL}
    \dl L = E_{ab} \dl g^{ab} + E_{A} \dl \phi^A + d\theta[\dl g, \dl \phi^A],
\end{equation}
where $E_{ab}$ and $E_{A}$ are proportional to the gravitational and matter equations of motion respectively, and $d\theta$ are total derivatives (including the variation of the boundary term $\ell$). Note that both the left and right hand sides of eq.\ \eqref{eq:dlL} are interpreted as one-forms on $\mc{C}$. The \emph{covariant phase space} $\mc{P}$ is then defined as the submanifold of $\mc{C}$ satisfying the equations of motion $E_{ab} = 0$ and $E_A = 0$. We further assume that there exists a pull-back $F_*$, that maps forms in $\mc{C}$ to forms in $\mc{P}$. In pedestrian terms, one should interpret this pullback as an operation that imposes both the equations of of motion, and their linearizations. Following the conventions of \cite{Bhattacharya:2025cld}, the wedge product between spacetime forms will be denoted by $\wedge$, whereas the wedge product between configuration space forms will be denoted by $\cw$.

The covariant symplectic current $\omega$ is a 2-form on $\mc{P}$ and a $(D-1)$-form on spacetime defined as
\begin{equation}
    \omega = F_*(\dl \theta).
\end{equation}
The associated symplectic form is the integral of $\omega$ over a Cauchy slice $\Sigma$:
\begin{equation}
    \Omega_{\Sigma} = \int_{\Sigma} \omega.
\end{equation}
It is easy to check that $\omega$ is a closed form, both in $\mc{C}$ and in spacetime. Consequently, $\Omega_{\Sigma}$ is independent of the choice of Cauchy slice $\Sigma$. However, $\Omega_{\Sigma}$ is degenerate because of gauge invariance.\footnote{For this reason $\mc{P}$ is often referred to as the \emph{pre phase space} in the literature. Strictly speaking, the true phase space of the theory is the symplectic quotient of $\mc{P}$ via the gauge group, which in this case is the group of ``small'' diffeomorphisms that vanish sufficiently fast at asymptotic infinity. This restores the non-degeneracy of $\Omega_{\Sigma}$. See \cite{Lee:1990nz, Harlow:2019yfa} for further discussion of issues pertaining to gauge invariance.} Let $\zeta$ be a vector field that generates a diffeomorphism of spacetime. We define a corresponding vector field $V_{\zeta}$ on $\mc{P}$ as
\begin{equation}
    V_{\zeta} = \int_M dx \left( \lie_{\zeta} g_{ab}(x) \frac{\dl}{\dl g_{ab}(x)} + \lie_{\zeta} \phi^A(x) \frac{\dl}{\dl \phi^A(x)} \right).
\end{equation}
If $\zeta$ is a ``small'' diffeomorphism that vanishes sufficiently fast at asymptotic infinity, then gauge invariance of the theory under $\zeta$ implies that the interior product of $\Omega_{\Sigma}$ with $V_{\zeta}$ vanishes. If, however, $\zeta$ is a ``large'' diffeomorphism that asymptotically approaches a Killing vector, then $I_{V_\zeta} \Omega_{\Sigma}$ will be non-zero. In fact, one can show that in \emph{any} diffeomorphism covariant theory of gravity, there exists a spacetime $(D-2)$-form $Q_{\zeta}$ such that
\begin{align} \label{eq:HWlocal}
    - I_{V_\zeta} \dl \theta &= d \chi_{\zeta} + G_{\zeta}, \\ \qquad \chi_{\zeta} &= \dl Q_{\zeta} - i_{\zeta} \theta, \\
    G_{\zeta} &= \zeta^a \dl (E_{ab} \epsilon^b) + \zeta^c \epsilon_c E_{ab} \dl g^{ab}.
\end{align}
Note that $G_{\zeta}$ is proportional to the equations of motion, and therefore vanishes once we pull back to phase space. Then, integrating the above on a Cauchy slice $\Sigma$ we find
\begin{equation} 
    -I_{V_\zeta} \Omega_{\Sigma} = \dl H_{\Sigma}(\zeta), \qquad H_{\Sigma}(\zeta) = \int_{\partial \Sigma} \chi_{\zeta}.
\end{equation}
This identity, known as the \emph{Hollands-Iyer-Wald} (HIW) identity, is simply the statement that asymptotic Killing vectors generate physical symmetries of gravity via codimension-2 charges. In Einstein gravity, one can show that the boundary charge $H_{\Sigma}(\zeta)$ can be written in terms of the Brown-York stress tensor as
\begin{equation} \label{eq:HBY}
    H_{\Sigma}(\zeta) = \int_{\partial \Sigma} \epsilon_a T^{ab}_{\text{BY}} \zeta_b, \qquad T^{ab}_{\text{BY}} = - \frac{1}{8 \pi G_N} \left( K^{ab} - h^{ab} K \right),
\end{equation}
where $K_{ab}$ is the extrinsic curvature of $\partial M$ with respect to the outward pointing normal, $h_{ab}$ is the induced metric on $\partial M$, and $K = h^{ab} K_{ab}$. In AdS/CFT, the Brown-York stress tensor, appropriately renormalized, can be identified with the CFT stress tensor \cite{Balasubramanian:1999re}. Therefore, in this case the HIW identity reduces to the familiar statement that the asymptotic symmetry charges of AdS generate the conformal group.

For concreteness, we present the exact expressions of the gravitational and matter symplectic forms for Einstein gravity minimally coupled to a massive scalar field:\footnote{The extra factors of 2 are there to compensate for the factors of $\frac{1}{2}$ coming from the wedge product $\cw$.}
\begin{align}
    \omega_{\text{matter}}|_{(g,\phi)} &=  - 2 (\dl \phi \cw \nabla^a \dl \phi)\,\eps_a - 2 \nabla_a \phi (\dl \phi \cw \dl g^{ab})\, \epsilon_b + g_{cd} \nabla^a \phi (\dl \phi \cw \dl g^{cd})\, \epsilon_a, \label{eq:Wmatter} \\
    \omega_{\text{grav}}|_{(g,\phi)} &= \frac{1}{8\pi G_N} P^{abcdef}(\dl g_{bc} \cw \nb_d \dl g_{ef} )\,\eps_a, \label{eq:Wgrav} \\
    P^{abcdef} &= g^{ae}g^{bf}g^{cd} - \frac{1}{2}g^{ad}g^{be}g^{cf} - \frac{1}{2} g^{ab}g^{cd}g^{ef} - \frac{1}{2} g^{ae}g^{bc}g^{fd} + \frac{1}{2} g^{ad}g^{bc}g^{ef}.
\end{align}
Note that on a fixed background without metric perturbations, the matter symplectic form reduces to the familiar expression $\omega_{\text{matter}} = 2 \epsilon_a \dl \pi^a \cw \dl \phi$.

The HIW identity also has important implications for gauge invariantly defined subregions of spacetime. For instance, let $X$ be a classically extremal (co-dimension 2) surface homologous to some boundary subregion $R$, and let $r$ be a (co-dimension 1) homology surface between $X$ and the boundary subregion $R$, such that $\partial r = X \cup R$. Now consider small perturbations $(\delta g_{ab},\delta \phi)$  of the metric and matter fields. One subtlety in this case is that the extremal surface $X$ itself will fluctuate when the metric fluctuates. It is therefore convenient to adopt a gauge where the coordinate location of the extremal surface remains unchanged as the metric changes (at least to leading order in perturbation theory). This gauge, often referred to as the \emph{Hollands-Wald gauge}, is defined by the following two conditions \cite{Hollands:2012sf}:
\begin{enumerate}
    \item \label{itm:HW1} The linearized deformation of the extremal surface $X^\1$ lies at the same coordinate location as the undeformed extremal surface $X^\0$.
    \item \label{itm:HW2} The boost vector field $\xi$ continues to satisfy the Killing equation in the deformed metric at the extremal surface, \emph{i.e.} $\lie_{\xi} ( g^\0 + g^\1 )|_{X^\0} = 0$.
\end{enumerate}
Then, in this gauge the identity eq.\ \eqref{eq:HWlocal} integrated on the subregion becomes
\begin{equation}
    -I_{V_\xi} \Omega_r  = \int_{R} \left( \dl Q_{\xi} - i_{\xi} \theta \right) - \int_{X^\0} \left(\dl Q_{\xi} - i_\xi \theta \right).
\end{equation}
One can show that the term localized at the extremal surface $X^\0$ is precisely the variation of the Wald entropy which is proportional to the area in Einstein gravity \cite{Wald:1993nt}. On the other hand, the term localized at the asymptotic boundary can be expressed as an integral of the Brown-York stress tensor as in eq.\ \eqref{eq:HBY}, and is equal to the variation of the expectation value of the CFT modular Hamiltonian \cite{Faulkner:2013ica}. Therefore, in the Hollands-Wald gauge we get
\begin{equation}
    -I_{V_\xi} \Omega_r  = \dl \langle K^R_{\text{CFT}} \rangle  - \frac{\dl \mc{A}(X^\0)}{4 G_N},
\end{equation}
where $\dl \langle K^R_{CFT} \rangle$ is the change in the CFT modular Hamiltonian and $\dl \mc{A}(X^\0)$ is the change in the area of the extremal surface. At first order in perturbation theory, this expression along with the CFT first law of entanglement, implies the RT formula \cite{Faulkner:2013ica}. At second order, it implies the duality between the boundary relative entropy and the gravitational canonical energy at large $N$ \cite{Lashkari:2015hha}. We will generalize these results to semi-classical gravity in section \ref{sec:REGrav}.


\subsection{Semi-classical gravity and the subregion symplectic form} \label{sec:HWSC}
In this subsection, we will review the generalization of the covariant phase space formalism to the case of semi-classical gravity, following \cite{Bhattacharya:2025cld}. In the presence of matter, the gravitational symplectic form is not conserved on its own. However, if matter is classical, then the sum of the gravitational and matter symplectic forms is indeed conserved. The main subtlety in semi-classical gravity is that matter resides in a quantum state, and therefore it is not immediately clear how to assign a symplectic form to the matter degrees of freedom. If the quantum state of matter is $\psi$, it is straightforward to check that
\begin{equation}
    d \omega_{\grav} = -\frac{1}{2} \dl \langle T_{ab} \rangle_{\psi} \cw \dl g^{ab},
\end{equation}
which implies that $\Omega^{\text{grav}}_{\Sigma}$ depends on the choice of Cauchy slice. It was argued in \cite{Bhattacharya:2025cld} that the appropriate quantity to add to the gravitational symplectic form that cancels the stress tensor contribution above is the Berry curvature of matter states viewed as a function of boundary sources. The sum of the gravitational symplectic form and the matter Berry curvature is independent of the choice of Cauchy slice and satisfies a semi-classical generalization of the HIW identity. Moreover, this construction has the correct classical limit since the Berry curvature of coherent states reproduces the classical symplectic form.

Concretely, we consider a family $\mc{P}_q$ of solutions $(g_{ab}(\lambda^i), \psi(\lambda^i))$ of the semi-classical Einstein equation. One can think of the space $\mc{P}_q$ as a semi-classical generalization of the classical covariant phase space $\mc{P}$. We can think of the parameters $\lambda^i$ as characterizing the boundary sources inserted in the boundary CFT Euclidean path integral state, which then serve as asymptotic boundary conditions for bulk fields.\footnote{It is known that the full space of solutions of the semi-classical Einstein equation is ill-behaved. There exist runaway solutions with exponentially growing back-reaction for small perturbations of the matter state around the vacuum \cite{Horowitz:1980fj}. Here we only consider the sub-family of well-behaved solutions.} We define the Berry connection for this family of states as
\begin{equation}
    \ba_{\Sigma} = - i \langle \psi | \dl \psi \rangle,
\end{equation}
where $\dl = \sum_i d\lambda^i \partial_{\lambda^i}$ acts on $(g_{ab}, \psi) \in \mc{P}_q$ by varying the boundary sources. Let $\bF_{\Sigma} = \dl \ba_{\Sigma}$ be the curvature associated to the connection $\ba_{\Sigma}$. We define the semi-classical symplectic 2-form as
\begin{equation}
    \bbF_{\Sigma} = \Omega^{\grav}_{\Sigma} + \bF_{\Sigma}.
\end{equation}
Then, one can check that $\bbF_{\Sigma} = \bbF_{\Sigma'}$ for any pair of Cauchy slices $\Sigma$ and $\Sigma'$. Now, given a vector field $\zeta$ on $M$, we define a vector field $V_{\zeta}$ that acts on $(g_{ab}, \psi)$ as follows:
\begin{align}
    I_{V_\zeta} \dl g_{ab} &= \lie_{\zeta} g_{ab} \\
    I_{V_\zeta} \delta | \psi \rangle &= -i \int_{\Sigma} \epsilon^a \zeta^b T_{ab} | \psi \rangle.
\end{align}
Then the semi-classical symplectic 2-form satisfies the following generalization of the classical HIW identity:
\begin{equation} \label{eq:HWSC}
    -I_{V_\zeta} \bbF_{\Sigma} = \int_{\partial \Sigma} \chi_{\zeta} + \int_{\Sigma} G^{\text{SC}}_{\zeta}.
\end{equation}
Here $\chi_{\zeta}$ is defined in eq.\ \eqref{eq:HWlocal}, and $G_{\zeta}$ is given by
\begin{equation} \label{eq:GSC}
    G_{\zeta}^{\text{SC}} =  \zeta^a \dl (E_{ab}^{\text{SC}} \epsilon^b) + \zeta^c \epsilon_c E_{ab}^{\text{SC}} \delta g^{ab}, 
\end{equation}
where we have defined
\begin{align}
    E_{ab}^{\text{SC}} = \frac{1}{16\pi G_N} \left( R_{ab} - \frac{1}{2} R g_{ab} + \Lambda g_{ab} - 8 \pi G_N \langle \psi | T_{ab} | \psi \rangle \right).
\end{align}
Just as in the classical case, once we pull back to the space of on-shell configurations $\mc{P}_q$, $G_{\zeta}^{\text{SC}}$ vanishes and we get
\begin{equation} \label{eq:HWSCGlobal}
    - I_{V_\zeta} \bbF_{\Sigma} = \dl H_{\Sigma}(\zeta), \qquad \dl H_{\Sigma}(\zeta) = \int_{\partial \Sigma} \chi_{\zeta}.
\end{equation}
The above identity is valid for global states defined on a complete Cauchy slice of $M$. 

We can also generalize eq.\ \eqref{eq:HWSCGlobal} to subregions in perturbation theory around a fixed background solution. Here it seems natural to look for a subregion notion of Berry connection. Note that any observer localized within $D(r)$ only has access to the reduced density matrix $\rho_{\psi} = \Tr_{\rbar} |\psi\rangle \langle \psi |$. To define a subregion Berry connection associated to the local state $\rho_{\psi}$, we construct a purification $\hpsi$ such that $\Tr_{r} |\hpsi\rangle\langle\hpsi| = \rho'_{\psi^\0}$, i.e., the density matrix on $\overline{r}$ looks like the vacuum corresponding to the background state $\psi_{(0)}$. We then use this purification to construct the subregion Berry connection in the usual way. As before, for the subregion case we need to work in the Hollands-Wald gauge which needs to be appropriately generalized since the bifurcation surface is now quantum extremal. We define the \emph{quantum} Hollands-Wald gauge via the following two conditions:
\begin{enumerate}
    \item \label{itm:QHW1} The linearized deformation of the QES $X^\1$ lies at the same coordinate location as the undeformed QES $X^\0$.
    \item \label{itm:QHW2} The boost vector field $\xi$ continues to satisfy the Killing equation in the deformed metric \emph{at} the QES.
\end{enumerate}
To assign a symplectic form to $D(r)$ we make one further assumption about the matter state $\psi$ --- we demand that the restriction of $\psi$ to $\bar{r}$ is close to the vacuum at very late times:
\begin{equation}
\lim_{s\to \infty}\langle \psi | \Delta_{\psi^{(0)}}^{-is} a' \Delta_{\psi^{(0)}}^{is} |\psi\rangle = \langle \psi^{(0)}|a' |\psi^{(0)}\rangle,\;\;\;a' \in \mathcal{M}_{\bar{r}}.
\end{equation}
Roughly speaking, we only consider states where everything either falls into the black hole or escapes to infinity, so that the local state in the exterior is indistinguishable from the vacuum state in the strict infinite time limit. Given such a state $\psi$, we define a new state $\hpsi$ as follows:
\begin{equation}
    |\hpsi \rangle = \lim_{s\to \infty} u'_{\psi^\0 | \psi}(s) | \psi \rangle,
\end{equation}
where $\psi^\0$ is the bulk vacuum state and $u'_{\psi^\0 | \psi}$ is the Connes cocycle associated to the complementary algebra $\mc{M}'$. Roughly speaking, the cocycle enacts an infinite boost on the left wedge which converts the state on $\bar{r}$ to the vacuum state \cite{Lashkari:2019ixo}.  In other words, for all operators $a \in \mc{M}$ and $b' \in \mc{M}'$ and for all states $\psi$ satisfying the stationarity condition, the purification $\hpsi$ satisfies
\begin{align} 
    \langle \hpsi | a | \hpsi \rangle &= \langle \psi | a | \psi \rangle, \label{eq:CCSew1}\\
    \langle \hpsi | b' | \hpsi \rangle &= \langle \psi^\0 | b' | \psi^\0 \rangle. \label{eq:CCSew2}
\end{align}
We then define the subregion Berry connection as
\begin{equation}
    \ba_r = -i \lim_{s\to\infty} \langle \hpsi_s | \dl \hpsi_s \rangle,
\end{equation}
and the associated Berry curvature as $\bF_r = \dl \ba_r$. The subregion semi-classical symplectic 2-form is then defined as
\begin{equation}
    \bbF_r = \Omega^{\text{grav}}_r + \bF_r.
\end{equation}
Now, let $\zeta$ be a vector field on $M$. Then the corresponding vector field $V_{\zeta}$ on $\mc{P}_q$ acts on the metric $g$ and the state $\hpsi$ as
\begin{align}
    I_{V_\zeta} \dl g_{ab} &= \lie_{\zeta} g_{ab}, \\
    I_{V_{\zeta}}\delta |\widehat{\psi} \rangle &= -i\int_{\Sigma} \epsilon_a  T^{ab} \zeta_b |\widehat{\psi} \rangle. \label{eq:IVpsi}
\end{align}
Similar to the global case, $\bbF_r$ is independent of the choice of homology surface $r$ and satisfies a subregion HIW identity:
\begin{equation} \label{eq:HWSCr}
    -I_{V_\zeta} \bbF_r = \int_{R} \chi_{\zeta} - \int_{X} \chi_{\zeta} + \int_{r} G_{\zeta}^{\text{SC}}.
\end{equation}
The key assumption that goes into the proof of the above equation are eqs.\ \eqref{eq:CCSew1} and \eqref{eq:CCSew2}. Roughly speaking, the fact that that the restriction of $\hpsi$ to the left wedge $D(\rbar)$ is the vacuum kills the part of eq.\ \eqref{eq:HWSC} supported on $\rbar$ and localizes the semiclassical symplectic two-form to the subregion $r$. Then, once the semi-classical Einstein equations are imposed, eq.\ \eqref{eq:HWSCsub} becomes
\begin{equation} \label{eq:HWSCsub}
    -I_{V_\zeta}\bbF_r = \dl \langle K^R_{\text{CFT}} \rangle - \frac{\dl \mc{A}(X^\0)}{4 G_N}.
\end{equation}
We emphasize that the above equation is valid only in the Hollands-Wald gauge perturbatively around a fixed background metric $g^\0$ and matter state $\psi^\0$.

\subsection{Perturbation theory for entanglement entropy}\label{sec:REGrav}
In this section we will study the entanglement entropy in perturbation theory up to $O(\lambda^2)$ using the quantum RT formula. To this end, we will use the semi-classical HIW identity presented in the previous section. The idea is to perturbatively expand eq.\ \eqref{eq:HWSCsub} around a fixed solution $(g^\0_{ab}, \psi^\0)$. Following the conventions of section \ref{sec:QES}, we will fix the background metric to be the AdS Rindler metric and the state $\psi^\0$ to be the vacuum state of matter fields. At first order, we will recover the Faulkner-Lewkowycz-Maldacena (FLM) formula \cite{Faulkner:2013ana}, and at second order we will generalize the results of \cite{Lashkari:2015hha}.  

Let $\xi$ be the bulk boost generator supported on the two wedges $D(r) \cup D(\rbar)$. Let $(g_{ab}(\lambda), \psi(\lambda))$ be a one-parameter family of solutions to the semi-classical Einstein equation dual to the boundary double trace deformed state, such that $(g_{ab}, \psi)|_{\lambda=0} = (g_{ab}^\0, \psi^\0)$. Then using eqs.\ \eqref{eq:CCSew1} and \eqref{eq:CCSew2} one can show that the matter part of the semi-classical symplectic two-form becomes a local integral of the expectation value of the bulk stress-tensor: 
\begin{equation} \label{eq:IVfT}
    -I_{V_\xi} \bF_r = \int_r \epsilon^a \xi^b \dl \langle T_{ab} \rangle_{\psi}.
\end{equation}
As mentioned before, although $\xi$ is supported on $D(r) \cup D(\rbar)$, this term is integrated only on $r$ since the infinite time limit of the cocycle sews the the matter state with the vacuum on $\rbar$, which removes the part supported on $\rbar$ (see \cite{Bhattacharya:2025cld} for a proof). Also note that eq.\ \eqref{eq:IVfT} is a purely kinematical identity that follows from the definitions \eqref{eq:IVpsi}, \eqref{eq:CCSew1} and \eqref{eq:CCSew2}. It is therefore valid for every value of $\lambda$. However, there might be additional subtleties when one considers gravitational effects beyond $O(G_N)$ since the Hollands-Wald gauge, as defined in section \ref{sec:HWSC}, fixes the location of the extremal surface only up to $O(G_N)$. In this paper we only consider $O(G_N)$ deformations of the extremal surface and ignore higher order effects.

Let us now expand the CFT state up to $O(\lambda)$. Since the deformation is due to a double trace operator, on the bulk side this leads to an $O(\lambda)$ change in the matter state which back-reacts on the geometry order by order in $G_N$ via the semi-classical Einstein equation. We thus expand the bulk solution $(g_{ab}, \psi)$ as follows:
\begin{align}
    \label{eq:SCexp}
    \begin{split}
        g_{ab} &= g^\0_{ab} + g^\1_{ab} + g^\2_{ab} + \cdots, \\
        \psi &= \psi^\0 + \psi^\1 + \psi^\2 + \cdots.
    \end{split}
\end{align}
Here $\psi^\1$ is the $O(\lambda)$ change in the bulk matter state due to the change in the boundary state (sans $1/N$ corrections), and $g^\1$ is the $O(\lambda G_N)$ change in the metric sourced by $\psi^\1$ through the linearized Einstein equation. Likewise, $\psi^\2$ is the $O(\lambda^2 G_N)$ change in the bulk matter state due to the change in the bulk metric, and $g^\2$ is the $O(\lambda^2 G_N^2)$ change in the metric sourced by $\psi^\2$ through the second order Einstein equation. The dots denote terms that are higher order in $G_N$. 

\subsubsection{First order}
At $O(\lambda),$ we do not need to account for the displacement of the QES since $X^\0$ is classically extremal. Thus the change in the area is only due to the change in the background metric. Moreover, recall that $\lie_{\xi} g^\0 = 0$, and therefore the gravitational symplectic form term drops out of eq.\ \eqref{eq:HWSCsub}. Then, using eq.\ \eqref{eq:IVfT} we get
\begin{equation}
    \langle K_{\text{CFT}} \rangle^\1 = \frac{\mc{A}^\1}{4 G_N} + \langle K_{\text{bulk}} \rangle^\1,
\end{equation}
where $\langle K_{\text{bulk}} \rangle^\1 = \int_r \epsilon^a \xi^b \langle T_{ab} \rangle^\1$ is the linearized change in the bulk modular Hamiltonian. This, along with the first law of entanglement implies that
\begin{equation}
    S^\1_{\text{CFT}} = \frac{\mc{A}^\1}{4 G_N} + S^\1_{\text{bulk}},
\end{equation}
which is precisely the FLM formula \cite{Faulkner:2013ana}.

\subsubsection{Second order}
At $O(\lambda^2)$ the extremal surface does fluctuate, so it is necessary to impose the Hollands-Wald gauge. Then, subtracting the change in the bulk entanglement entropy from both sides of equation \eqref{eq:HWSCsub} we get
\begin{equation}
    \langle K_{\text{CFT}} \rangle^\2 - S_{\text{gen}}^\2 = -(I_{V_\xi} \Omega^{\text{grav}}_r)^\2 + \left( \int_r \xi^a \left( \langle T^{ab} \rangle \epsilon_b\right)^\2 - S_{\text{bulk}}^\2 \right).
\end{equation}
Identifying $S^\2_{\text{gen}}$ with $S^\2_{\text{CFT}}$ and using the identity $S(\psi \| \omega) = \dl \langle K \rangle - \dl S$ we get
\begin{equation} \label{eq:SrelEq}
    S^\2_{\text{CFT}} (\Psi \| \Omega) = \Omega^{\text{grav}}_{r}(g^\1, \lie_{\xi} g^\1) + \left( \int_r \xi^a \left( \langle T^{ab} \rangle \epsilon_b\right)^\2 - S_{\text{bulk}}^\2 \right),
\end{equation}
where we have explicitly written out the gravitational symplectic form in components. This result is consistent with those of \cite{Jafferis:2015del, Lashkari:2015hha}. Indeed, for a single trace deformation of the CFT state, $\psi$ is a coherent state. Therefore $\dl S_{\text{bulk}} = 0$\footnote{This is because the reduced density matrix of a coherent state is unitarily equivalent to the reduced density matrix of the vacuum state: $\rho_r^{\text{coherent}} = U_r \rho_r^\0 U_r^\dagger$ where $U_r = e^{ i \int_r (\pi \widehat{\phi} - \phi \widehat{\pi})}$.} and by the arguments of section \ref{sec:HWSC}, the stress tensor term reduces to the matter symplectic form on $r$. This is precisely what was observed in \cite{Lashkari:2015hha}. Here we propose that eq.\ \eqref{eq:SrelEq} is the correct generalization of the results of \cite{Lashkari:2015hha} for non-coherent matter states arising from general multi-trace deformations of the CFT vacuum. Note that the terms in parentheses in eq.\ \eqref{eq:SrelEq} can be expressed as the matter relative entropy $S_{\text{rel}}(\psi \| \psi^\0)$ plus additional corrections due to backreaction. Our results are therefore consistent with those of \cite{Dong:2017xht}, where it was argued that the bulk and boundary relative entropies are not expected to match beyond $O(1)$ in the $G_N$ expansion.

\subsubsection{Moving away from the Hollands-Wald gauge}
Our goal is to reproduce eq.\ \eqref{eq:SrelEq} from a first principles CFT computation assuming nothing but the standard AdS/CFT dictionary \cite{Witten:1998qj, Gubser:1998bc}. One caveat is that eq.\ \eqref{eq:SrelEq} is valid only in the Hollands-Wald gauge, whereas the CFT computation will be more convenient in a more physical gauge, such as the transverse traceless gauge. In order to compare the two computations, it would be useful to express eq.\ \eqref{eq:SrelEq} in an arbitrary gauge.

Let $g^\1_{ab}$ denote the metric perturbation in the Hollands-Wald gauge and let $h^\1_{ab}$ be the metric perturbation in some arbitrary gauge. Let $\mc{V}$ be an $O(\lambda G_N)$ vector field that moves us to the Hollands-Wald gauge from the new gauge:
\begin{equation} \label{eq:GTM}
    g^\1 = h^\1 + \lie_{\mc{V}} g^\0,
\end{equation}
The vector field $\mc{V}$ is often referred to as the Hollands-Wald vector field. Note that $\mc{V}$ is precisely the displacement profile of the QES, and therefore it is sufficient to take $\mc{V}$ to be supported in a small neighborhood of the QES (see appendix \ref{app:HW} for further details). Under this transformation, the semi-classical symplectic form changes by a boundary term. To see this note that $\bbF_r$ is anti-symmetric and bilinear in the metric, and thus
\begin{align}
    \begin{split}
        \bbF_r(g^\1, \lie_\xi g^\1) &= \bbF_r(h^\1 + \lie_\mc{V} g^\0, \lie_\xi(h^\1 + \lie_{\mc{V}} g^\0)) \\
        &= \bbF_r(h^\1, \lie_\xi h^\1) + \int_{X} \chi_{[\xi, \mc{V}]}(h^\1) - \int_{X} \chi_\mc{V}(\lie_\xi(h^\1 +\lie_\mc{V} g^\0)).
    \end{split}
\end{align}
where, for notational simplicity we have suppressed the dependence of $\bbF_r$ on the state. In the second line we have used the linearized semi-classical HIW identity and fact that $\xi$ is a background Killing vector field to write $\lie_{\xi} \lie_{\mc{V}} g^\0 = \lie_{[\xi, \mc{V}]} g^\0$. We have also thrown away boundary terms at infinity since we may assume that $\mc{V}$ decays to zero sufficiently fast near infinity. Now, it can be shown by direct computation that $\chi_{\mc{V}}(\lie_\xi(h + \lie_{\mc{V}} g^\0))|_{X} = 0$ (see \cite{Faulkner:2017tkh}). Then we are left with only one boundary term localized at $X$:
\begin{equation}
    \bbF_r (g^\1, \lie_{\xi} g^\1) = \bbF_r(h^\1, \lie_{\xi} h^\1) - \int_X \chi_{[\xi, \mc{V}]}(h^\1)
\end{equation}
Splitting up the semi-classical symplectic form in to a gravitational term and a matter term, and plugging back into eq.\ \eqref{eq:SrelEq} we get
\begin{equation} \label{eq:SrelMain}
    S^\2_{\text{CFT}}(\Psi \| \Omega) = \Omega^{\text{grav}}(h^\1, \lie_{\xi} h^\1) -  \int_{X} \chi_{[\xi, \mc{V}]}(h^\1) + \cdots
\end{equation}
where the $+ \cdots$ denotes the additional matter terms coming from eq.\ \eqref{eq:SrelEq}. In the following sections we will see how to reproduce the two gravitational terms shown above from a CFT calculation. In the process, we will show that consistency of the bulk and boundary computations requires $\mc{V}$ to satisfy the displacement profile formula eq.\ \eqref{eq:vplus}. Of course, the full CFT relative entropy also contains the additional matter terms that we have suppressed. We will subsequently ignore them and leave reproducing the full expression for the relative entropy for future work.

\section{CFT relative entropy from Modular Witten diagrams} \label{sec:SK}

The goal of this section is to now compute the entanglement entropy from the CFT side to $O(\lambda^2)$, and then use bulk Witten diagrams to compute to $O(\lambda G_N)$ and reproduce aspects of the quantum extremality formula. At second order in the CFT, the first law of entanglement gets replaced by the more general statement:
\begin{equation}
    \Delta S_R^{\Psi} = \Delta K_R - S_{\text{rel}}(\rho_R^{\Psi}|| \rho_R^\0).
\end{equation}
We see by comparison with equation \eqref{eq:SrelEq} that this would match with the gravity calculation if the CFT relative entropy at $O(\lambda^2)$ agrees with the canonical energy plus the bulk relative entropy in the correctly deformed entanglement wedge from the bulk side. To this end, we will compute the boundary relative entropy at $O(\lambda^2)$ directly in the CFT, and show its equivalence to the bulk calculation assuming the standard AdS/CFT dictionary for correlation functions. Our approach closely follows that of \cite{Faulkner:2017tkh}, with the important difference that our state deformation involves double-trace operators, and so we will have to deal with more complicated Witten diagrams in the bulk to get $O(\lambda^2 G_N)$ effects. 
\subsection{Setup}
Recall that the CFT state $\Psi$ is defined by turning on an $O(1)$ source for a double-trace operator $O^{(2)}$ in the Euclidean path integral. Such Euclidean path integral states are natural to consider from the AdS/CFT point of view \cite{Marolf:2017kvq}. Concretely, let $\phi$ be the elementary fields in our path integral, and $\varphi(\mathbf{x})$ be the corresponding boundary data specified on the $x^0_E = 0$ slice. The state $\ket{\Psi(\lambda)}$ is defined by the following Euclidean path integral:
\begin{equation} 
    \braket{\varphi(\mathbf{x}) | \Psi(\lambda)} = \int^{\phi(0, \mathbf{x}) = \varphi(\mathbf{x})} D\phi\ \exp \left[ -\int_{-\infty}^0 dx^0_E \int d^{d-1}\mathbf{x} (\mc{L}_{\text{CFT}} + \lambda\,J(x^0_E, \mathbf{x}) O^\2(x^0_E, \mathbf{x})) \right].
\end{equation}
Clearly at $\lambda=0$, the state $\ket{\Psi(0)}$ reduces to the vacuum state. For normalizability, we require that source smoothly turns off as $x^0_E \to 0$. Now, let $\rho_R(\lambda)$ be the corresponding reduced density operator for the state $\Psi(\lambda)$ on the subregion $R$. The perturbative expansion for the reduced density operator takes the form \cite{Rosenhaus:2014woa}:
\begin{equation} \label{eq:DMPert}
    \rho_R(\lambda) = \rho_R^\0 + \lambda \int_{0}^{2\pi} d\tau\,\int d^{d-1} \mathbf{y}\, \widetilde{J}(\tau, \mathbf{y})\, \rho_R^\0  O^\2(\tau, \mathbf{y})+ \mc{O}(\lambda^2),
\end{equation}
where $\tau$ is the angular coordinate around the entanglement cut (whose analytic continuation to Lorentz signature gives the boost angle), $\mathbf{y}$ denotes spatial coordinates on the subregion $A$, and we have defined the time-reflection symmetric source:
\begin{equation}
    \widetilde{J}(\tau, \mathbf{y}) = \begin{cases}
        J(\tau, \mathbf{y}), \quad \text{when } -\pi < \tau < 0 \\
        J^*(2\pi-\tau, \mathbf{y}), \quad \text{when } 0 < \tau < \pi
    \end{cases}
\end{equation}
Further, $\rho_R^\0 = e^{-K_R^{(0)}}$ is the reduced density operator associated to the vacuum state. Recall that for the region $R$ being a half space, the vacuum modular Hamiltonian $K_R^\0 = -\log \rho_R^\0$ is the half-sided boost operator.

The relative entropy between the states $\rho_A(\lambda)$ and $\rho_A^\0$ is given by:
\begin{equation}
S_{\text{rel}}(\rho_R(\lambda)|| \rho_R^0) =\mathrm{Tr}\left(\rho_R(\lambda)\log\,\rho_R(\lambda)-\rho_R(\lambda)\log \,\rho_R^\0\right) .
\end{equation}
Differentiating the relative entropy with respect to $\lambda$, we get
\begin{eqnarray}
    \frac{d}{d\lambda} S_{\text{CFT}}(\rho_R(\lambda) \| \rho_R^\0) &=& \Tr \left[ \frac{d \rho_R}{d\lambda} \log \rho_R + \rho_R \frac{d}{d\lambda} \log \rho_R - \frac{d\rho_R}{d\lambda} \log \rho_R^\0 \right]\nonumber\\
    &=& \Tr \left[ \frac{d \rho_R}{d\lambda} \left(\log \rho_R  -  \log \rho_R^\0 \right)\right],
\end{eqnarray}
where we have used $\mathrm{Tr}\left(\frac{d}{d\lambda}\rho_R(\lambda)\right)=0$. Of course, evaluating the derivative at $\lambda=0$ yields the first law of entanglement,
\begin{equation}
    \frac{d}{d\lambda} S_{\text{CFT}}(\rho_R \| \rho_R^\0)\Big|_{\lambda=0} = 0.
\end{equation}
Therefore, the leading non-zero contribution is at $O(\lambda^2)$. Differentiating once again and evaluating the result at $\lambda=0$, we find
\begin{equation} \label{eq:d2Srel}
    S^\2_{\text{CFT}}(\rho_R \| \rho_R^\0) \equiv \frac{d^2}{d\lambda^2} S_{\text{CFT}}(\rho_R \| \rho_R^\0)\Big|_{\lambda=0} = -\Tr\left[ \dl \rho_R \dl K_R \right],
\end{equation}
where $\dl \rho_R = (d\rho_R/d\lambda)_{\lambda=0}$ and $\dl K_R = -(d \log \rho_R/d\lambda)_{\lambda=0}$ are the $O(\lambda)$ changes in the density matrix and the modular Hamiltonian respectively. The perturbative correction to the modular Hamiltonian at first order is given by \cite{Faulkner:2016mzt, Balakrishnan:2020lbp}
\begin{equation} \label{eq:d1KA}
    \dl K_R = \int d\tau \int d^{d-1}\mathbf{y}\, \widetilde{J}(\tau,\mathbf{y})\int_{-\infty}^{\infty} \frac{ds}{4\sinh^2 \left(\frac{s+i\tau}{2} \right)} O^{(2)}(s,\mathbf{y}),
\end{equation}
where note that
$$ O^{(2)}(s,\mathbf{y}) = e^{isK_R^\0/2\pi} O^{(2)}(\tau=0,\mathbf{y}) e^{-isK_R^\0/2\pi}$$
is a Lorentzian operator, with $K_R^\0 = -\log \rho_R^\0$ being the modular Hamiltonian of the unperturbed state, i.e., the half-sided boost operator. Using equations \eqref{eq:DMPert} and \eqref{eq:d1KA}, we thus find\footnote{While we have motivated this formula in a setting where the Hilbert space factorizes, the formula is actually more general and can be derived even for Type III algebras using Tomita-Takesaki theory \cite{toappear}.}
\begin{equation}
    S_{\text{CFT}}^\2(\rho_R \| \rho_R^\0) = - \int d\mu\, \int_{-\infty}^{\infty } \frac{ds}{f(s+i\tau_2)} \langle O^\2(\tau_1, \mathbf{y}_1) O^\2(s, \mathbf{y}_2) \rangle,
\end{equation}
where we have defined
\begin{equation}
    \int d\mu = \int d\tau_1 \int d^{d-1} \mathbf{y}_1 \int d\tau_2 \int d^{d-1}\mathbf{y}_2 \, \widetilde{J}(\tau_1, \mathbf{y}_1)\, \widetilde{J}(\tau_2, \mathbf{y}_2),
\end{equation}
and $f(s)$ is the function:
\begin{equation}
    f(s) = 4\sinh^2\left( \frac{s}{2} \right).
\end{equation}
At this stage, it convenient to perform a conformal transformation to $S^1 \times \HH^{d-1}$ \cite{Casini:2011kv}. One can show that the modular flow in these new coordinates corresponds to a translation in the $S^1$ direction.  Denoting coordinates on $\HH^{d-1}$ with the symbol $Y$, our expression for the relative entropy now becomes
\begin{equation} \label{eq:Srel_hyperbolic}
    S_{\text{CFT}}^\2(\rho_R \| \rho_R^\0) = -\int d\boldsymbol{\mu} \int_{-\infty}^{\infty}\frac{ds}{f(s + i\tau_2)} \langle O^\2(\tau_1, Y_1) O^\2(s, Y_2) \rangle
\end{equation}
where we have absorbed the appropriate conformal factors inside the new measure $d\boldsymbol{\mu}$. 

In order for the CFT calculation to match with the quantum RT formula calculation, we need to show that the above formula for the relative entropy on the CFT side agrees with equation \eqref{eq:SrelEq}, including the correct deformation of the quantum extremal surface. Note that two-point correction function of double-trace operators appearing above is slightly non-standard, in that it involves modular flow. While Euclidean correlation functions can be computed using standard Witten diagrammatic rules in AdS/CFT, it is not immediately clear how to use these rules in the case of modular-flowed correlation functions. Below, we will have to develop some understanding of Witten diagrams for modular-flowed correlation functions.

\subsection{Modular Witten diagrams} \label{sec:invHKLL}
Using boundary time translation invariance, the two-point function defined above can be written as 
\begin{equation} \label{2ptFcases}
\langle O(\tau,Y_1)O(s,Y_2)\rangle = \begin{cases} \langle \Psi(\tau_1,Y_1) | \Psi(s,Y_2)\rangle &\cdots \;\;0 < \tau_1 < \pi,\\ \langle \Psi(s,Y_2) | \Psi(2\pi -\tau_1,Y_1)\rangle &\cdots\;\; \pi < \tau_1 <2\pi \end{cases}
\end{equation}
where $|\Psi(\tau,Y)\rangle$ is defined as the Euclidean path integral with the operator $O$ inserted at Euclidean modular time $-\tau$ in the lower half plane, and
\begin{equation}
|\Psi(s,Y_2)\rangle = O(s,Y_2) |\Omega\rangle.
\end{equation}
where $\ket{\Omega}$ is the CFT vacuum state. Our goal is to re-write this state from the bulk point of view using Witten diagrams, and then to use this to obtain a bulk expression for the above two-point function. 

\subsubsection*{Warm-up: Scalar single-trace operators} \label{sec:Single-trace}
As a warm-up, let us first work through the simpler case where our boundary operator insertion is a single trace operator.  Consider the state $\ket{\Psi(s, Y_2)} = O(s, Y_2) |\Omega\rangle$. Let $\phi$ be the bulk scalar field dual the single-trace operator $O$. In the large-$N$ limit, we can treat the bulk scalar as a free field, and the CFT vacuum $|\Omega\rangle$ as the Fock vacuum $|0\rangle$ from the bulk point of view. Let $\Sigma$ be the time-reflection symmetric Cauchy surface in the bulk. We can write $\phi$ as:
\begin{equation}
    \phi(\sigma) = \sum_k (f_k(\sigma) a_k + f^*_k(\sigma) a^\dagger_k),
\end{equation}
where $\sigma \in \Sigma$ and $\{f_k\}$ are a complete set of solutions to the Klein-Gordon equation in the Lorentzian section. Recall that the Klein-Gordon inner product between two such solutions on $\Sigma$ is defined as
\begin{equation}
    \braket{f, g} = i\int_\Sigma \omega(f, g) = i\int_{\Sigma}\left[ (n^a\pa_{a}f)\,g - f \,(n^{a}\pa_{a}g)\right],
\end{equation}
where $n^{a}$ is the time-like normal vector field to $\Sigma$. We will choose the mode functions to satisfy $\braket{f^*_k, f_{k^\prime}} = \dl_{kk^\prime}$.

Consider the case $0 < \tau_1 < \pi$. We can insert a complete set of bulk single particle states inside the two-point function \eqref{2ptFcases} and rewrite it as follows:
\begin{align}
    \begin{split}
        \langle \Psi(\tau_1,Y_1) | \Psi(s,Y_2)\rangle &= \sum_k \braket{\Psi(\tau_1,Y_1) |  a_k^\dagger | 0} \braket{0|a_k O(s, Y_2) | \Omega} + \cdots\\
        &= \sum_{k, k^\prime} \braket{\Psi(\tau_1,Y_1)| a_k^\dagger | 0} \braket{0|a_{k^\prime} O(s, Y_2) | \Omega} \delta_{kk^\prime} + \cdots \\
        &= \sum_{k, k^\prime} \braket{\Psi(\tau_1,Y_1) |  a_k^\dagger | 0} \braket{0|a_{k^\prime} O(s, Y_2) | \Omega} i \int_{\Sigma} \omega\left( f^*_k, f_{k^\prime} \right) + \cdots
    \end{split}
\end{align}
where $+ \cdots$ denotes multiparticle states which contribute at higher orders in $1/N$. In the second line above, we have introduced a second sum over momenta together with a delta function, and in the third line we have re-written this delta function in terms of the KG inner product of two mode functions. Using linearity of the symplectic flux, the two-point function can then be written as
\begin{equation}
    \langle \Psi(\tau_1,Y_1) | \Psi(s,Y_2)\rangle = \int_{\Sigma} \omega \left( \braket{\Psi(\tau_1,Y_1)| \sum_k f^*_k a^\dagger_k | 0 }, i\braket{0 | \sum_{k^\prime} f_{k^\prime} a_{k^\prime} O(s, Y_2) | \Omega} \right).
\end{equation}
Now, at leading order in $1/N$ the bulk Fock space vacuum $\ket{0}$ and the CFT vacuum $\ket{\Omega}$ are interchangeable. This lets us rewrite the second argument of the symplectic flux as
\begin{equation}
    \sum_{k^\prime} \braket{0 |  f_{k^\prime} a_{k^\prime} O(s, Y_2) | \Omega} = \sum_{k^\prime} \braket{0 | [  f_{k^\prime} a_{k^\prime}, O(s, Y_2) ]| \Omega}
\end{equation}
since $a_{k^\prime} \ket{\Omega} = a_{k^\prime} \ket{0} = 0$. Further, we may replace the sum $\sum_{k^\prime} f_{k^\prime} a_{k^\prime}$ with $\phi = \sum_{k^\prime} (f_{k^\prime} a_{k^\prime} + f^*_{k^\prime} a^\dagger_{k^\prime})$. This is because the additional term contains an $f_{k^\prime}^*$ whose symplectic overlap with the first argument of the symplectic flux gives zero. Therefore, it follows that the second argument of the symplectic flux is precisely the causal bulk-to-boundary propagator defined as
\begin{equation}
    K_C(\sigma | s, Y_2) = i \braket{\Omega | [\phi(\sigma), O(s, Y_2)] | \Omega}, \quad \sigma \in \Sigma.
\end{equation}
Similarly, since $a_k$ annihilates $\ket{0}$, we can replace $\sum_k f^*_k a^\dagger_k$ with $\phi$ in the first argument of the symplectic flux. This shows that the first argument is precisely the Euclidean bulk-to-boundary propagator defined as
\begin{equation}
    K_E(\sigma | \tau_1, Y_1) = \braket{\Psi(\tau_1,Y_1) | \phi(\sigma) | \Omega}, \quad \sigma \in \Sigma.
\end{equation}
Collecting everything together, we see that the leading order contribution to the two-point function for single trace operators can be written as
\begin{equation}
    \langle O(\tau_1, Y_1)O(s,Y_2)\rangle = \int_{\Sigma} \omega\left(K_E(\sigma | \tau_1, Y_1), K_C(\sigma | s, Y_2)\right).
\end{equation}
The other case  where $ \pi < \tau_1 < 2\pi$ also works similarly and gives the same formula. What we have accomplished here is that we have re-written the boundary two point function as a manifestly bulk integral over the time-reflection symmetric slice $\Sigma$. It was shown in \cite{Faulkner:2017tkh} (and as we will see later in this paper), the integral over the modular time $s$ can now be performed efficiently in this form, and the result precisely agrees with the bulk canonical energy within the correctly deformed entanglement wedge.

\subsubsection*{Double trace operators}
Let us now turn to the double trace case. Our goal is to rewrite the double trace 2-point function as a bulk quantity using the standard rules of AdS/CFT for correlation functions, but in a particular way, such that the integral over modular time becomes easy to perform. For this, we will need to invoke a particular matter action in the bulk-dual to the boundary holographic CFT. For concreteness, we will assume that the bulk matter comprises of a single scalar field $\phi$, dual to a single trace operator $O$ in the CFT, although the details of the matter action will not be too important for our purposes. Then, turning on the corresponding double trace operator $O^{(2)} = :OO:$ in the CFT path integral deforms the quantum state of the bulk scalar. Importantly, we will assume that the bulk graviton is minimally coupled to the matter stress tensor:
\begin{equation}
    S_{\text{int}} = \int d^{d+1}x \sqrt{-g}\, T^{ab}h_{ab}.
\end{equation} 

\begin{figure}[t]
    \centering
    \includegraphics[width=0.6\linewidth]{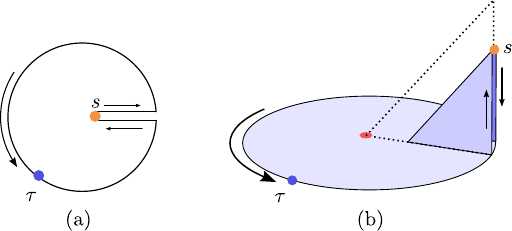}
    \caption{(a) The contour of integration of the CFT path integral. The blue dot indicates the Euclidean operator insertion $O^\2(\tau)$ and the orange dot indicates the Lorentzian insertion $O^\2(s)$. For simplicity we have suppressed the transverse directions. The arrows indicate the direction of operator ordering inside the path integral. (b) The bulk geometry dual to the boundary path integral. The blue and orange dots indicate the boundary operator insertions and the red dot at the center is the extremal surface in the background geometry. The dotted line denotes the full entanglement wedge dual to the boundary subregion in the background geometry. For Witten diagrams, the region of interest will be the shaded region which lies in the causal past of the orange point.}
    \label{fig:gr-sk}
\end{figure}

Note that in the CFT 2-point function that appears in eq.\ \eqref{eq:Srel_hyperbolic}, one operator insertion is the Euclidean section while the other one is in the Lorentzian section. The boundary path integral which computes such correlation functions is typically performed along a Schwinger-Keldysh (SK) contour in the complex $s$-plane with a single real time fold, as illustrated in figure \ref{fig:gr-sk}. The standard AdS/CFT dictionary for Euclidean correlation functions \cite{Witten:1998qj} instructs us to compute all Witten diagrams in the Euclidean bulk saddle geometry by pulling down bulk interaction vertices order by order. In order to compute the SK correlation function relevant to us, we need to suitably generalize this prescription to time-folds on the boundary. In the black hole context, this was pioneered in \cite{Glorioso:2018mmw, Jana:2020vyx}, where the resulting complex bulk geometry was called the gravitational Schwinger-Keldysh geometry. In the present context, we are in very similar setting because the boundary vacuum state is thermal with respect to the modular Hamiltonian of $R$, and correspondingly the bulk vacuum state is thermal with respect to the modular Hamiltonian for the AdS Rindler wedge.  So, a natural prescription to follow is: 
\begin{enumerate}
    \item think of the boundary operators as bulk operators using the extrapolate dictionary, 
    \item  replace boundary modular flow with bulk modular flow, and
    \item treat the resulting bulk modular-flowed correlation function as a thermal SK correlation function (from the bulk point of view) and use standard Feynman rules of thermal field theory in the bulk to compute. 
\end{enumerate}
We will call the resulting Witten diagrams \emph{modular Witten diagrams}. We will first begin by systematically laying out the Feynman rules for how to compute such modular Witten diagrams in the specific case of the double-trace two-point function in eq. \eqref{eq:Srel_hyperbolic}. In the next section, we will give a  justification for the above rules via analytic continuation from the purely Euclidean correlation function. Our main focus here will be on the $s$-channel graviton exchange diagram, which we will show can be re-written (using the above rules) in terms of the bulk gravitational symplectic form. 

A couple of remarks: in the single-trace case, we were essentially dealing with the bulk two-point function (appropriately extrapolated to the boundary), and so we did not need to worry too much about the SK contour; the two point function could simply be written down by analytic continuation from the Euclidean case. The main difference here is that the two-point function of double-trace operators (at the first subleading order in $G_N$) involves Witten diagrams in the bulk, which makes the analytic continuation slightly delicate, and we are forced to deal with bulk SK-contours. Secondly, we are in the fortunate situation where modular flow for the vacuum state is geometric. For more general background states and subregions, the problem seems more involved, but we expect the same set of Witten diagram rules to work.

\subsubsection*{Feynman rules for modular Schwinger-Keldysh}
Let us start with some definitions. We define the Euclidean bulk-to-boundary propagator for the bulk matter stress tensor as
\begin{align} \label{eq:bulk-to-bdry-eucl}
    \begin{split}
        K_{ab}( \tau_x, \tilde{x} | \tau, Y_1 )
        &= \langle T_{ab}(\tau_x, \tilde{x}) O^\2(\tau_1, Y_1) \rangle,
    \end{split}
\end{align}
where $\tilde{x}$ denotes all the other transverse coordinates in the bulk and $\langle \cdots \rangle$ stands for the Euclidean path integral with the specified insertions. The causal bulk-to-boundary propagator is defined via analytic continuation of the Euclidean propagator as
\begin{equation} \label{eq:bulk-to-bdry-causal}
    K_{C;ab}(s_x, \tilde{x} | s, Y_1 ) = i \lim_{\eps \rightarrow 0} \left[ K_{ab}(i(s_x + i\eps), \tilde{x} | is, Y_1) - K_{ab}(i(s_x - i\eps), \tilde{x} | is, \tilde{y}) \right],
\end{equation}
with the advanced/retarded propagators defined from the causal propagator in the usual way. We will also need the Euclidean bulk-to-bulk graviton propagator, which is defined as
\begin{align} \label{eq:bulk-to-bulk-eucl}
    \begin{split}
        G_{abcd}(\tau_x, \tilde{x} | \tau_y, \tilde{y}) &= \langle h_{ab}(\tau_x, \tilde{x}) h_{cd}(\tau_y, \tilde{y}) \rangle,
    \end{split}
\end{align}
while the causal bulk-to-bulk propagator is given by
\begin{align} \label{eq:bulk-to-bulk-causal}
    \begin{split}
        G_{C;abcd}(s_x, \tilde{x} | s_y, \tilde{y}) &= i \lim_{\eps \rightarrow 0} \left[ G_{abcd}(i(s_x + i\eps), \tilde{x} | is_y, \tilde{y}) - G_{abcd}(i(s_x - i\eps), \tilde{x} | is_y, \tilde{y}) \right] \\
        &= i \langle 0 | [h_{ab}(s_x, \tilde{x}), h_{cd}(s_y, \tilde{y})] | 0 \rangle
    \end{split}
\end{align}
where $|0\rangle$ is the bulk Fock space vacuum. As before, advanced/retarded bulk-to-bulk propagators are defined from the causal propagator in the standard way. In what follows, the graviton propagators will be written in the transverse, traceless gauge:
\begin{equation}
{h^a}_a = 0, \;\; \nabla_a{h^a}_b=0.
\end{equation}
This gauge has the advantage that all the graviton bulk to boundary and bulk to bulk propagators do not have any acausal (or spacelike) branch cuts \cite{Costa:2014kfa}. In other gauges, the propagators may develop un-physical cuts as gauge-artifacts, but the transverse, traceless gauge avoids these issues. (Note however, that in comparing with the gravity calculation, we will have to go back to the generalized Hollands-Wald gauge discussed previously.)

As mentioned previously, the natural prescription to compute SK correlators in the CFT is to perform the bulk path integral along the corresponding SK contour. All operator insertions in the path integral will thus be contour ordered, as indicated by the arrows in figure \ref{fig:gr-sk}. Note that the Lorentzian section of the geometry contains two sheets, one corresponding to forward and the other corresponding to backward time evolution. If an interaction vertex is inserted in the Lorentzian section, then its contraction with the Lorentzian boundary point $s$ yields a causal propagator (i.e. a commutator) due to the combination of forward and backward evolutions, and the contour ordering prescription. This is not true if the Lorentzian vertex is contracted with a point in the Euclidean section of the geometry. So, we get the following Feynman rules for Witten diagrams (see appendix \ref{app:Feyn-rules} for further discussion):
\begin{enumerate}
    \item All propagators connecting two Euclidean vertices or a Euclidean vertex with a Lorentzian vertex are Wightman.\footnote{In the literature such propagators are often called Schwinger functions, whereas the name Wightman function is reserved for real time $n$-point correlators. However, it is well known that Schwinger functions and Wightman functions are related by analytic continuation (e.g. see chapter 19 of \cite{Glimm:1987ylb}). In this paper we will refer to all 2-point correlators in the complexified geometry as Wightman.}
    \item All propagators connecting two Lorentzian vertices are causal. They may be either retarded or advanced depending on the sign of the boundary time coordinate $s$.
\end{enumerate}
These Feynman rules are very standard in the context of SK correlation functions in thermal field theory (see, for instance \cite{Stanford:2015owe}, where these rules were used for a perturbative computation out-of-time-ordered correlators in quantum field theory), and the claim here is that we must simply adopt these rules in the bulk to compute boundary SK correlation functions. In the rest of this section, we will follow this prescription to compute the SK two point function for double-trace operators in terms of bulk Witten diagrams. We will then justify the above prescription from analytic continuation in the next section. 

\begin{figure}[t]
    \centering
    \includegraphics[width=0.7\linewidth]{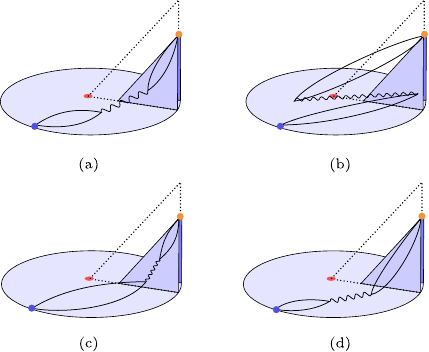}
    \caption{The $s$-channel graviton exchange diagrams. Different figures correspond to vertex insertions at different locations in the geometry. (a) One vertex is Euclidean and the other is Lorentzian. (b) One vertex is Euclidean and the other is Loretzian, but the contractions are ``criss-crossed''. (c) Both vertices are Lorentzian. (d) Both vertices are Euclidean.}
    \label{fig:s-graviton}
\end{figure}

\subsubsection*{Witten diagrams on the bulk SK contour}
Let us first focus on the $s$-channel graviton exchange diagram. In this diagram, the two interaction vertices can be inserted at various locations along the contour, as shown in figure \ref{fig:s-graviton}. Note that a Lorentzian interaction vertex contracted with the Lorentzian boundary insertion at $s$ need only be integrated over the causal past of $s$; this is because all the propagators connecting two points on the Lorentzian part of the contour are necessarily causal propagators, as per the Feynman rules stated previously. The only exception to this rule is diagram (b) --- but it is easy to check that this diagram vanishes identically since the forward and backward evolutions exactly cancel out. Throughout the rest of this paper, we will only focus of diagram (a), with one vertex inserted in the Euclidean section and the other in the Lorentzian section. The corresponding Witten diagram can be written as:
\begin{equation}
    \int_0^{2\pi} d\tau_x \int d\tilde{x} \int_0^s ds_y \int d\tilde{y}\, K^{ab}(\tau_x, \tilde{x} | \tau_1, Y_1)\, G_{abcd}(\tau_x, \tilde{x} | s_y, \tilde{y})\, K_{C}^{cd}(s_y, \tilde{y} | s, Y_2).
\end{equation}
Then, we cut open the bulk-to-bulk graviton propagator on the time-reflection symmetric slice $\Sigma$ by inserting a complete set of single graviton states and perform the same set of manipulations shown in the single trace case. Concretely, we first mode expand the graviton operator on $\Sigma$ as:
\begin{equation}
    h_{pq}(\sigma) = \sum_{k} \left( f_{k;pq}(\sigma) b_k + f^*_{k;pq}(\sigma) b^\dagger_k \right),
\end{equation}
where $b_k$ denotes the graviton annihilation operator, and the mode functions $f_{k;pq}$ form a complete set of solutions of the linearized Einstein equation in the Lorentzian section. As before, the mode functions are normalized with respect to the symplectic inner product:
\begin{equation}
    \langle f_{k}, f_{k^\prime}\rangle = \int_\Sigma \omega_{\text{grav}}(f^*_{k;pq}, f_{k^\prime;rs}) = \delta_{kk^\prime},
\end{equation}
where $\omega_{\text{grav}}$ is the gravitational symplectic current defined in eq.\ \eqref{eq:Wgrav}. Then, inserting a complete set of single graviton states in the bulk-to-bulk propagator $G_{abcd}$ we obtain:
\begin{align}
    \begin{split}
        G_{abcd}(\tau_x, \tilde{x} | s_y, \tilde{y}) &= \langle   h_{ab}(\tau_x, \tilde{x}) |h_{cd}(s_y, \tilde{y}) | 0\rangle \\ 
        &= i\sum_{k, k^\prime} \langle   h_{ab}(\tau_x, \tilde{x}) |b^\dagger_k | 0\rangle \langle 0 | b_{k^\prime} h_{cd}(s_y, \tilde{y} | 0\rangle \int_\Sigma \omega_{\text{grav}}\left(f^*_{k;pq}(\sigma) , f_{k^\prime;rs}(\sigma)\right) \\
        &= \int_\Sigma \omega_{\text{grav}} \left( \langle h_{ab}(\tau_x, \tilde{x}) |\sum_k f^*_{k;pq}(\sigma) b^\dagger_k | 0\rangle, i \langle 0 | [ \sum_{k^\prime} f_{k^\prime;rs}(\sigma) b_{k^\prime},  h_{cd}(s_y, \tilde{y}) ] |0  \rangle\right) \\
        &= \int_\Sigma \omega_{\text{grav}} \left(G_{abpq}(\tau_x, \tilde{x} | \sigma), G_{C;rscd}(\sigma | s_y, \tilde{y})  \right),
    \end{split}
\end{align}
where, recall that $G_{C;rscd}$ is the causal bulk-to-bulk propagator. Next, we push the bulk-to-boundary propagators inside the symplectic form and obtain
\begin{equation} \label{eq:1_graviton_Symp}
    \int_\Sigma \omega_{\text{grav}} \left( \mc{K}_{pq}(\sigma | \tau, Y_1), \mc{K}_{C;rs}(\sigma | s, Y_2)  \right).
\end{equation}
where we have defined
\begin{align}
    \mc{K}_{pq}(\sigma | \tau, Y_1) &= \int_0^{2\pi} d\tau_x \int d\tilde{x}\, K^{ab}(\tau_x, \tilde{x} | \tau, Y_1)\, G_{abpq}(\tau_x, \tilde{x} | \sigma) \label{eq:Kdef} \\
    \mc{K}_{C;rs}(\sigma | s, Y_2) &= \int_0^s ds_y \int d\tilde{y}\, K^{cd}_C(s_y, \tilde{y})\, G_{C;cdrs}(\sigma | s_y, \tilde{y}). \label{eq:KCdef}
\end{align}
Note that it follows from the Feynman rules stated previously that (see appendix \ref{app:Feyn-rules} for further details)
\begin{equation}
    \mc{K}_{ab} = \langle h_{ab}(\sigma)O^{(2)}(\tau_1,Y_1)\rangle,
\end{equation}
\begin{equation}
    \mc{K}_{C;cd} = \langle \Omega| [h_{cd}(\sigma),O^{(2)}(s,Y_2)]|\Omega\rangle,
\end{equation}
are equal to the above Wightman and causal two-point functions respectively (see Appendix \ref{app:Feyn-rules}). In summary, we have managed to write one of the contributions (fig. \ref{fig:s-graviton}, panel (a)) to the $s$-channel graviton exchange in terms of the gravitational symplectic form in the bulk. We will see that performing the modular $s$-integral in this term will lead us to the gravitational part of the bulk answer for relative entropy. 


While in this work we will focus on the diagram isolated above, we expect that all the other diagrams, either $s$-channel or $t$-channel, will contribute only to the bulk relative entropy term. For instance, one can systematically get rid of the $t$-channel diagrams by considering a large $N_F$ number of flavors of the scalar operator and choosing to deform by the operator $O^{(2)} \sim \frac{1}{N_F}\sum_{I=1}^{N_F} :O_I O_I:$. This naturally has the effect of suppressing $t$-channel diagrams. As for the remaining $s$-channel diagrams in figure \ref{fig:s-graviton}, the diagrams in (c) and (d) roughly arise because turning on the double-trace operator causes back-reaction, which then further modifies the quantum state of the bulk matter fields (because it is defined via the Euclidean path integral on the said deformed metric). Diagram (b) of course simply vanishes. In what follows, we will focus on the contribution isolated in equation \eqref{eq:1_graviton_Symp} and show that we can recover the gravitational symplectic flux, including the correct boundary terms corresponding to the gauge transformation between the transverse, traceless gauge and the Hollands-Wald gauge. But before that, we take a slight detour to show how the Feynman rules we used above come out directly from analytic continuation of Euclidean correlation functions.

\subsection{Detour: Modular Witten diagrams from analytic continuation} \label{sec:grsk-derivation}
\begin{figure}[t]
    \centering
    \includegraphics[width=0.9\linewidth]{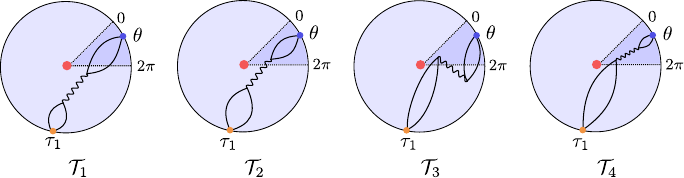}
    \caption{The 4 diagrams corresponding to the 4 terms in eq. \eqref{eq:eucl-split}. The region in light blue is the regular region lying between $0$ and $2\pi$. The shaded region is the excess lying between $2\pi$ and $2\pi\alpha$. Note that $\tau=0$ is identified with $\tau = 2\pi\alpha$. }
    \label{fig:2-pt-eucl}
\end{figure}

We will focus on the specific case of the two point function of double-trace operators. We start with the Euclidean two-point function:
\begin{align}
    \begin{split}
       G^{(\alpha)}_{O^\2 O^\2}(\tau_1, Y_1; \theta, Y_2) &= \langle O(\tau_1, Y_1)  O^\2(\theta, Y_2) \rangle_\alpha
    \end{split}
\end{align}
in a Euclidean bulk geometry at a slightly higher temperature $2\pi\alpha$, with $\alpha > 1$. The first operator is inserted in the region $0<\tau_1<2\pi$, while the second operator is inserted in the region $2\pi < \theta < 2\pi\alpha$. We will refer to the regions $(0,2\pi)$ as the regular region and $(2\pi, 2\pi\alpha)$ as the excess region. These are shown in figure \ref{fig:2-pt-eucl}. We will use standard rules of AdS/CFT to compute this Euclidean two-point function using Witten diagrams. The goal will be to analytically continue $\theta$ to real time, and then take the limit $\alpha \rightarrow 1$, so that one lands on the desired Schwinger-Keldysh correlation function at the original temperature. 

At leading order, we pull down two interaction vertices with the corresponding $s$-channel Witten diagram being\footnote{Since we are interested in one of the contributions coming from the $s$-channel, we only present the argument here for this case. We expect the $t$-channel diagrams to work similarly.}
\begin{equation} \label{eq:2-pt-Eucl}
    G^{(\alpha)}_{O^\2 O^\2}(\tau;\theta) = \int_0^{2\pi\alpha}d\tau_x \int_0^{2\pi\alpha} d\tau_y\, K^{ab}_\alpha(\tau_x | \tau) G^\alpha_{abcd}(\tau_x | \tau_y) K^{cd}_{\alpha}(\tau_y | \theta),
\end{equation}
where, for notational simplicity we have suppressed the dependence on the spatial coordinates and displayed only the Euclidean time integrals. Note that the bulk-to-bulk and bulk-to-boundary propagators implicitly depend on $\alpha$ through the bulk geometry, as indicated by the sub/superscripts. In the limit $\alpha \rightarrow 1$ these reduce to the usual bulk-to-bulk and bulk-to-boundary propagators defined in eqs. \eqref{eq:bulk-to-bulk-eucl} and \eqref{eq:bulk-to-bdry-eucl} respectively.

Next, we split up the $\tau_x$ and $\tau_y$ integrals into regular and excess parts. In other words, we rewrite eq. \eqref{eq:2-pt-Eucl} as
\begin{align} \label{eq:eucl-split}
    \begin{split}
        G_{O^\2 O^\2}(\tau;\theta) &= \int_0^{2\pi}d\tau_x \int_0^{2\pi} d\tau_y\, K^{ab}_\alpha(\tau_x | \tau) G^\alpha_{abcd}(\tau_x | \tau_y) K^{cd}_{\alpha}(\tau_y | \theta) \\
        &+ \int_0^{2\pi} d\tau_x \int_{2\pi}^{2\pi\alpha}d\tau_y\, K^{ab}_\alpha(\tau_x | \tau) G^\alpha_{abcd}(\tau_x | \tau_y) K^{cd}_{\alpha}(\tau_y | \theta) \\
        &+ \int_{2\pi}^{2\pi\alpha}d\tau_x \int_0^{2\pi} d\tau_y\, K^{ab}_\alpha(\tau_x | \tau) G^\alpha_{abcd}(\tau_x | \tau_y) K^{cd}_{\alpha}(\tau_y | \theta) \\
        &+ \int_{2\pi}^{2\pi\alpha}d\tau_x \int_{2\pi}^{2\pi\alpha} d\tau_y\, K^{ab}_\alpha(\tau_x | \tau) G^\alpha_{abcd}(\tau_x | \tau_y) K^{cd}_{\alpha}(\tau_y | \theta) \\
        &:= \mc{T}_1 + \mc{T}_2 + \mc{T}_3 + \mc{T}_4.
    \end{split}
\end{align}
These 4 terms are pictured in figure \ref{fig:2-pt-eucl}. We want to analytically continue $\theta \rightarrow is$. But in doing so, we must deform the $\tau_x$ and $\tau_y$ contour integrals above so as to make sure that we do not cross any singularities/ branch cuts; this will naturally lead us to the SK contour in the bulk. We will now process the 4 terms one by one.

\begin{figure}[t]
    \centering
    \includegraphics[width=\linewidth]{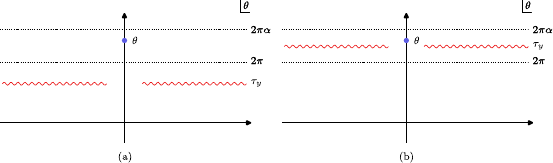}
    \caption{(a) Analytic structure of $\mc{T}_1$ in the complex $\theta$ plane where both $\tau_x, \tau_y \in (0, 2\pi)$. The branch cut corresponding to $\tau_y$ is denoted by the squiggly red line. (b) Analytic structure of $\mc{T}_2$ where the $\tau_y$ branch cut lies in the excess region $(2\pi, 2\pi\alpha)$.}
    \label{fig:theta-contour}
\end{figure}

First let us focus on the first term $\mathcal{T}_1$ in eq. \eqref{eq:eucl-split} and look at its analytic structure in the complex $\theta$ plane. Here $\theta$ lies in the excess region whereas $\tau_x$ and $\tau_y$ lie in the regular region. Therefore the function is analytic in the strip $\mc{S}$ defined as $\mc{S} := \{ z \in \mathbb{C}\, |\, \operatorname{Im} z \in (2\pi, 2\pi\alpha) \}$ as shown in the left panel of figure \ref{fig:theta-contour}. Therefore we can simply continue $\theta \rightarrow 2\pi + is$ without issue and take the limit $\alpha \rightarrow 1$. But there are subtleties when one of the branch cuts lies in $\mc{S}$.

\begin{figure}
    \centering
    \includegraphics[width=\linewidth]{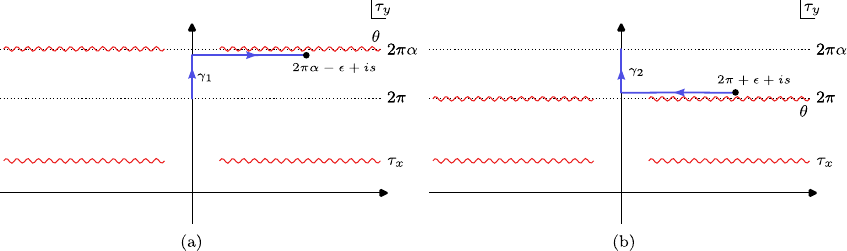}
    \caption{Analytic continuation of the bulk point $\tau_y$. The branch cut corresponding to $\tau_x$ comes from the bulk-to-bulk propagator, while the cut corresponding to $\theta$ comes from the bulk-to-boundary propagator. (a) Integration contour $\gamma_1$ when $\theta > \tau_y$. (b) Integration contour $\gamma_2$ when $\theta < \tau_y$.}
    \label{fig:bulk-cont}
\end{figure}

In the second term both $\theta$ and $\tau_y$ lie in the excess region, as shown in the right panel of figure \ref{fig:theta-contour}. Here the idea is to continue $\theta$ in such a way so that the $\tau_y$ branch cut is never crossed, \emph{i.e.} if $\theta > \tau_y$ we continue $\theta \rightarrow 2\pi\alpha + is$ and if $\theta < \tau_y$ we continue $\theta \rightarrow 2\pi + is$. These might seem like two different analytic continuations, but in the limit $\alpha \rightarrow 1$ they will coincide. To process $\mc{T}_2$ we now split the $\tau_y$ integral into two parts $\tau_y < \theta$ and $\tau_y > \theta$ and perform the analytic continuation accordingly. Therefore
\begin{align}
    \begin{split}
        \mc{T}_2 = \int_0^{2\pi} d\tau_x\, K_\alpha^{ab}(\tau_x | \tau)\left[ \int_{2\pi}^\theta d\tau_y\, G^\alpha_{abcd}(\tau_x | \tau_y) K^{cd}_\alpha(\tau_y | \theta) + \int_{\theta}^{2\pi\alpha} d\tau_y\, G^\alpha_{abcd}(\tau_x | \tau_y) K^{cd}_\alpha(\tau_y | \theta) \right].
    \end{split}
\end{align}
As discussed above, in the first term we continue $\theta \rightarrow 2\pi\alpha + is$ and in the second term we continue $\theta \rightarrow 2\pi + is$. Then the two terms in square brackets above become
\begin{equation}
    \int_{\gamma_1} d\tau_y\, G^{\alpha}_{abcd}(\tau_x | \tau_y) K_{\alpha}^{cd}(\tau_y | 2\pi\alpha + is) + \int_{\gamma_2} d\tau_y\, G^{\alpha}_{abcd}(\tau_x | \tau_y) K_{\alpha}^{cd}(\tau_y | 2\pi + is)
\end{equation}
where $\gamma_1$ and $\gamma_2$ are the contours shown in figure \ref{fig:bulk-cont}. If we now look at the analytic structure of the integrand in the complex $\tau_y$ plane, we observe that in the first term the branch cut corresponding to $\theta$ is lifted to $2\pi\alpha + is$ and in the second term the branch cut descends to $2\pi + is$. Now that the branch cuts have moved to the boundaries of the strip $\mc{S}$, $\gamma_1$ and $\gamma_2$ can be chosen to be any contour with the appropriate endpoints since the correlator is analytic in $\mc{S}$. At this stage, we take the limit $\alpha \rightarrow 1$ so that the height of the strip $\mc{S}$ goes to zero. In this limit, the contributions coming from the vertical segments of $\gamma_1$ and $\gamma_2$ drop out and we only pick up the discontinuity across the the $\theta$ branch cut. Therefore, as $\alpha \rightarrow 1$, $\mc{T}_2$ becomes
\begin{align}
    \begin{split}
        \mc{T}_2 &= \int_0^{2\pi} d\tau_x\, K^{ab}(\tau_x | \tau) \int_0^s ds_y \operatorname{Disc} \left[G_{abcd}(\tau_x | is_y) K^{cd}(s_y | is) \right] \\
        &= \int_0^{2\pi} d\tau_x\, K^{ab}(\tau_x | \tau) \int_0^s ds_y\, G_{abcd}(\tau_x | is) K^{cd}_C(s_y | s)
    \end{split}
\end{align}
which is precisely the graviton exchange diagram with one vertex insertion in the Euclidean section and the other in the Lorentzian section.

The other two diagrams can also be processed analogously. Performing similar manipulations, one can see that $\mc{T}_3$ vanishes. This is consistent with the fact that diagram (b) in figure \ref{fig:s-graviton} also vanishes. Finally, $\mc{T}_4$ corresponds to diagram (c) of figure \ref{fig:s-graviton} with both interaction vertices inserted in the Lorentzian segment. We will not analyze these diagrams here. The interested reader can find the relevant details in appendix \ref{app:T3T4}.


It is worth noting that the locality of modular flow was used crucially in the above discussion; this is what allowed us to change the temperature to $2\pi \alpha$ and take the limit $\alpha \to 1$. In the more general case without local modular flow, one must resort to the replica trick \cite{Faulkner:2018faa}. Given any subregion $A$ in a QFT with an associated reduced density matrix $\rho_A$, the modular flow replica trick gives us a recipe to compute the following 2-point function:
\begin{equation}
    \langle O(\tau, x) O(is, y) \rangle := \Tr \left[ \rho_A O(\tau,x) \rho_A^{-is}O(0, y) \rho_A^{is} \right].
\end{equation}
The idea is to start with the purely Euclidean two-point function
\begin{equation} \label{eq:2-pt-repl}
    \langle O(\tau, x) O(0,y) \rangle_{n,k} := \Tr \left[ \rho_A^n O(\tau, x) \rho_A^{-k} O(0, y) \rho_A^{k} \right],
\end{equation}
and analytically continue $n \rightarrow 1$ and $k \rightarrow is$. In the absence of local modular flow, the analytic continuation is not straightforward. However, if the theory is holographic, one can instead replace the right hand side of eq. \eqref{eq:2-pt-repl} with a bulk path integral on the geometry $\mc{B}_n$ dual to the boundary replica manifold. As shown in \cite{Lewkowycz:2013nqa}, the assumption of replica symmetry for the bulk saddle then allows one to perform the analytic continuation in the limit $n\to 1$. Subsequently, one can continue $k \rightarrow is$. We expect that this procedure gives a similar set of Feynman rules for bulk Witten diagrams, where now real time evolution is replaced with bulk modular flow. Of course, these diagrams cannot be ``visualized'' in the usual way because modular flow is a non-local operation in general, but one can still formally write down modular Witten diagrams. It would be interesting to work this out carefully, but we leave this for future work.

\subsection{Processing the Graviton Term} \label{sec:graviton}
\begin{figure}[t]
    \centering
    \includegraphics[width=\linewidth]{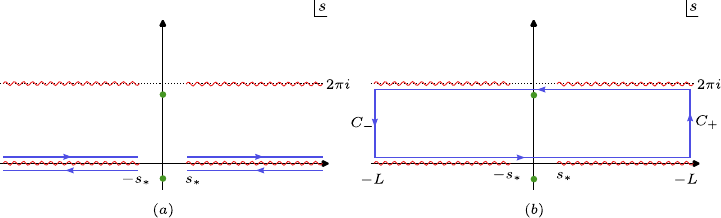}
    \caption{The strip $\mathrm{Im}(s) \in [0, 2\pi]$ in the complex plane. The red wavy lines correspond to the branch cuts of $K_{C;ab}$ while the dotted lines are the regions where $K_{C;ab}$ vanishes. The blue dots denote the poles of the modular kernel $1/f(s+i\epsilon)$ and the vertical contours are denoted by $C_{\pm}$.}
    \label{fig:contour}
\end{figure}

Let us now return to the analysis of the $s$-channel graviton exchange diagram. The rewriting of the graviton exchange diagram in terms of the bulk symplectic form (see eq.\ \eqref{eq:1_graviton_Symp}) is useful in that it allows us to compute the corresponding contribution to the boundary relative entropy by performing the integral over the modular flow parameter $s$. Plugging in eq.\ \eqref{eq:1_graviton_Symp} into eq.\ \eqref{eq:Srel_hyperbolic} we can rewrite the relative entropy as
\begin{equation} \label{eq:RESF}
    S^\2(\rho_A \| \rho^\0_A) = -\int d\boldsymbol{\mu} \int_{-\infty}^{\infty} \frac{ds}{f(s+i\tau_2)} \int_{\Sigma} \omega_{\text{grav}}\left( \mc{K}_{pq}(\sigma | \tau_1, Y_1), \mc{K}_{C;rs}(\sigma | s, Y_2) \right) + \cdots,
\end{equation}
where the $\cdots$ denote the other $s$-channel and $t$-channel diagrams that are not being considered here. Since the $s$-dependence comes only from the causal propagator, our main object of interest is the integral: 
\begin{equation} \label{eq:sint-main}
    I_{rs} = \int_{-\infty}^{\infty} \frac{ds}{f(s+i\eps)} \mc{K}_{C;rs}(\sigma|s, Y_2)
\end{equation}
\begin{figure}[t]
    \centering
    \includegraphics[height=6.8cm]{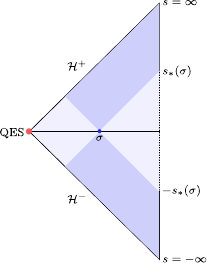}
    \caption{Support of the causal propagator $\mc{K}_{C;rs}$. Given a fixed $\sigma$, the shaded region denotes the location where $\mc{K}_{C;rs}$ has non-trivial support. Therefore the $s$-integral vanishes along the vertical dotted line from $-s_*(\sigma)$ to $s_*(\sigma)$.}
    \label{fig:CP}
\end{figure}
Note that the causal propagator $\mc{K}_{C;rs}(\sigma | s, Y_2) = \langle [h_{rs}(\sigma), O^\2(s, Y_2)]\rangle$  vanishes whenever the bulk point $\sigma$ is spacelike separated from the boundary point $(s, Y_2)$. For this reason, we can restrict the integral over $\Sigma$ to lie on the homology surface $r$ dual to the boundray subregion $R$. Furthermore, given any $\sigma \in r$, there exists some boundary time $s_*(\sigma)$ such that $\mc{K}_{C}(\sigma | s, Y_2)$ vanishes in the interval $-s_*(\sigma) < s < s_*(\sigma)$ (see figure \ref{fig:CP}). 
Now, it follows from definition that the causal propagator evaluates the discontinuity across a branch cut in the complex $s$-plane, which corresponds to the region where the operators are timelike separated. Therefore, we may write (see panel (a) of figure \ref{fig:contour})
\begin{equation}
    I_{rs} = i \int_{-\infty}^{\infty} \frac{ds}{f(s+i\tau_2)} \Disc\, \mc{K}_{rs}(\sigma | is, Y_2),
\end{equation}
where, recall, $\mc{K}_{rs} = \langle h_{rs} O^\2\rangle$ is the Euclidean bulk-to-boundary two-point function. Due to KMS periodicity of modular flowed correlators, the discontinuity repeats at intervals of $2\pi$ in the imaginary direction. Note that the discontinuity vanishes in the intermediate segment $(-s_*(\sigma), s_*(\sigma))$ on account of the operators being spacelike separated in this region. Then, using KMS periodicity, we lift the integral along the lower branch of the cut to $\RR + 2\pi i$, and close the contour in such a way that it encloses the pole at $s= -i\tau_2$ coming from the kernel $f(s)$, at the cost of adding and subtracting vertical contours at infinity. Let $C_{\pm}$ be the vertical contours at $\mathrm{Re}(s) = \pm \infty$, and $\Gamma$ be the closed contour lying in the strip $\mc{S}$ that we get by adding the vertical contributions at infinity, as shown in panel (b) of figure \ref{fig:contour}. Then we have
\begin{equation} \label{eq:s-int}
    I_{rs} = i \oint_{\Gamma} \frac{ds}{f(s+i\tau_2)}\mc{K}_{rs}(\sigma|is,Y) + \hat{I}_{rs}(C_{+}) + \hat{I}_{rs}(C_{-}).
\end{equation}
where $\hat{I}_{ab}(C_{\pm})$ are the vertical contour terms defined as
\begin{equation}
    \hat{I}_{rs}(C_{\pm}) = \mp i\lim_{L\rightarrow\infty}\int_{\pm L}^{\pm L+2\pi i} \frac{ds}{f(s + i\tau_2)}\,\mc{K}_{rs}(\sigma|is,Y).
\end{equation}
The contour integral over $\Gamma$ simply collects the residue at the pole. Looking ahead, this term will precisely correspond to the symplectic flux term in eq. \eqref{eq:SrelMain}. On the other hand, the vertical contour integrals will evaluate to a boundary term localized on the QES. Then demanding that this term match up with the boundary terms in \eqref{eq:SrelMain} will let us recover the quantum extremality formula for the shape of the QES. Let us now analyze these terms one by one.

\subsubsection*{Pole contribution}
For the integral over $\Gamma$, the residue from the double pole at $s=-i\tau_2$ gives
\begin{equation}
    i\oint_{\Gamma} \frac{ds}{4\sinh^2(\frac{s + i\tau_2}{2})} \mc{K}_{rs}(\sigma|is,Y) = -2\pi \frac{\partial}{\partial s} \mc{K}_{rs}(\sigma|\tau_2, Y).
\end{equation}
Then, using time translation invariance we can make the replacement $-2\pi\partial_s \rightarrow \lie_{\xi}$ and obtain 
\begin{equation}
  I_{rs} = \lie_{\xi} \mc{K}_{rs}(\sigma|0,Y) + \hat{I}_{rs}(C_{+}) + \hat{I}_{rs}(C_{-}).
\end{equation}
Therefore, the pole contribution to the relative entropy becomes
\begin{equation}
    S^\2(\rho_R \| \rho^\0_R)_{\text{pole}} = \int d\boldsymbol{\mu} \int_{\Sigma} \omega_{\text{grav}}\left( \mc{K}_{pq}(\sigma | \tau_1, Y_1), \lie_\xi \mc{K}_{rs}(\sigma|\tau_2,Y) \right).
\end{equation}
Plugging this into eq. \eqref{eq:RESF} and performing the Euclidean integrals in $d\boldsymbol{\mu}$, we arrive at
\begin{equation}
  S^\2(\rho_R \| \rho^\0_R)_{\text{pole}} = \int_\Sigma \omega_{\text{grav}}(\langle h\rangle, \lie_{\xi}\langle h\rangle),
\end{equation}
where $\langle h_{ab}\rangle$ is the bulk graviton one-point function. This precisely matches with the symplectic flux term in equation \eqref{eq:SrelMain} on the gravity side.

\subsubsection*{Vertical contours}
Next, let us process the vertical contour contributions at $\pm \infty$. First note that using translational invariance of the bulk-to-boundary two-point function, we may transfer the $s$ dependence to the bulk metric and write
\begin{equation}
    \begin{split}
        \mc{K}_{ab}(\sigma|is,Y) &= \langle h_{ab}(-s, \tilde{x}) O^\2(0, Y) \rangle \\
        &= \langle e^{-i s K_R^\0} h_{ab}(0, \tilde{x}) e^{is K_R^\0} O^\2(Y) \rangle
    \end{split}
\end{equation}
where $K^{\0}_R$ is the vacuum modular Hamiltonian on $R$. Plugging this into the $s$-integral we get 
\begin{equation}
    \hat{I}_{ab}(C_{\pm}) = \mp i \lim_{L\rightarrow \infty}\int_{\pm L}^{\pm L + 2\pi i} \frac{ds}{f(s + i\tau_2)} \langle e^{-is K_R^\0} h_{ab}(x^+, x^-, y) e^{is K_{R}^\0} O^\2(Y) \rangle,
\end{equation}
where we have evaluated the bulk metric on the $t=0$ slice, which, in light cone coordinates is given by $x^+ = x^-$. Note that boundary modular flow acts nicely on bulk local operators, i.e., as a local boost in the background spacetime. In particular, under modular flow the bulk metric transforms as
\begin{equation}
    e^{-isK^{\0}_A} h_{ab}(x^+, x^-, y) e^{isK^{\0}_A} = J\indices{^c_a}(-s) J\indices{^d_b}(-s) h_{cd}(e^{-s} x^+, e^s x^-, y),
\end{equation}
where $J$ is the boost matrix defined as
\begin{equation}
  J\indices{^M_N}(s) =
  \begin{pmatrix}
    e^s & 0 \\
    0 & e^{-s}
  \end{pmatrix} \qquad
  J\indices{^{\alpha}_{\beta}} = \delta\indices{^{\alpha}_{\beta}} \qquad J\indices{^M_{\alpha}} = J\indices{^{\alpha}_{M}} = 0.
\end{equation}
Thus, the vertical contour integral at $-\infty$ becomes:
\begin{align}
    \hat{I}_{ab}(C_-) = - \lim_{L \to -\infty}\int_0^{2\pi} \frac{d\theta}{f(L+i(\theta+\tau_2))}& J\indices{^c_a}(-L-i\theta)J\indices{^d_b}(-L-i\theta) \nonumber\\ & \times
    \langle h_{cd} ( e^{-L - i \theta} x^+, e^{L + i\theta} x^-, y) O^\2(Y) \rangle
\end{align}
In the $L\rightarrow -  \infty$ limit, the bulk point gets boosted off to infinity. Since we expect the correlation function above to decay exponentially in $L$, the integral above seems like it should naively vanish. However, this argument is too quick. 
Let us take a careful look at the individual components of $\hat{I}$ starting with $\hat{I}_{++}$. A short computation reveals that for large values of $L$ we have
\begin{equation}
  \hat{I}_{++}(C_-) \simeq - \int_0^{2\pi} d\theta\, e^{- L - i\theta} \langle h_{++}(e^{- L - i \theta} x^+, e^{L + i\theta} x^-, y) O^\2(Y) \rangle
\end{equation}
For $x^+ \neq 0$, it is clear that the bulk metric operator is getting boosted off to infinity as $L \to -\infty$, and so the correlation function will decay exponentially. However, when the bulk operator starts approaching $x^+ = 0$, the overall factor of $e^{-L}$ (coming from the boost acting on the indices of the metric operator) blows up. Thus, it is natural to guess that $\hat{I}_{++}$ is proportional to $\delta(x^+)$. To extract the coefficient of this delta function, we can smear $\hat{I}_{++}$ in the $x^+$ direction:\footnote{We can smear the operator in a more symmetric way $x^+\in (-\epsilon,\epsilon)$ if desired, but because of the relative minus sign in the boost between the two complementary Rindler wedges, the part of the operator in $(-\epsilon,0)$ does not contribute in the $L\to -\infty$ limit.}
\begin{equation} 
\int_{0}^{\epsilon} dx^+ \hat{I}_{++}.
\end{equation}
Changing the variable to $\tilde{x}^+ = e^{- L - i\theta} x^+$ we get
\begin{align}
  \int_{0}^{\epsilon} dx^+ \hat{I}_{++}(C_-) &= - \lim_{L\to -\infty} \int_0^{2\pi}d\theta \int_{0}^{\epsilon e^{-L} e^{-i \theta}} d\tilde{x}^+ \langle h_{++}(\tilde{x}^+, e^{L + i \theta} x^-, y) O^\2 \rangle \\
  &= \int_{0}^{2\pi}d\theta \int_{0}^{\infty e^{-i \theta}} d\tilde{x}^+ \langle h_{++}(\tilde{x}^{+}, 0, y) O^\2(Y) \rangle .
\end{align}
By analyticity of the two point function, we rotate the contour back to the real line. Then, the $\theta$ integral evaluates to an overall factor of $2\pi$, and we conclude that
\begin{equation} 
\hat{I}_{++}(C_-) = - 2\pi \int_{0}^{\infty} dx^+ \langle h_{++}(0, x^-, y) O^\2(Y_2)\rangle.
\end{equation}
Repeating this analysis for the other components of the metric reveals that they vanish in the $L\to -\infty$ limit. Similarly, for the contour at $s=-\infty$, the only non-vanishing component of $\hat{I}_{ab}$ is $\hat{I}_{--}$. 
Inserting this back into equation \eqref{eq:RESF} and integrating over the boundary points, we thus get
\begin{equation} \label{eq:REFinal}
    S^\2(\rho_A \| \rho_A^{\0}) = \Omega^{\text{grav}}_{r}(\langle h \rangle, \lie_{\xi}\langle h \rangle) + \Omega^{\text{grav}}_{r}( \langle h \rangle, \mc{I}_{++}(C_{-})) + \Omega_{r}^{\text{grav}}(\langle h \rangle, \mc{I}_{--}(C_+)) + \cdots
\end{equation}
where we have defined
\begin{align}
  \mc{I}_{++}(C_-) = -2\pi \delta(x^+) \int_0^{\infty} dx^{+} \langle h_{++}(x^+,0,y) \rangle,
\end{align}
and $\mc{I}_{--}$ can be defined analogously. Now recall that the bulk slice $\Sigma$ was the time reflection symmetric slice defined by the equation $x^+ = x^-$. Because of the $\delta$ functions in $\mc{I}_{\pm \pm}$ it is clear that the vertical contours evaluate to a boundary term localized at the quantum extremal surface $x^+ = x^{-} =0$. It is therefore reasonable to expect that the vertical contour contributions correspond to the $\int_{\tilde{A}}\chi$ term in eq.\ \eqref{eq:SrelMain}. 

To see this explicitly, recall from the discussion in section \ref{sec:HWSC} that the gravitational symplectic form satisfies the identity
\begin{equation} \label{eq:HWmatch}
    \Omega^{\text{grav}}_{r}(\langle h \rangle, \lie_{\zeta} g^\0) = - \int_{X}\chi_{\zeta}(\langle h \rangle) - \int_r \zeta^a \langle T_{ab} \rangle^\1 \epsilon^b,
\end{equation}
as long as $\langle h \rangle$ solves the semi-classical linearized Einstein equation, and $\zeta$ decays sufficiently fast at infinity. If we ignore the bulk term integrated on $r$ for the moment, comparing eqs.\ \eqref{eq:REFinal} and \eqref{eq:HWmatch} with the boundary term in \eqref{eq:SrelMain}, we conclude that it is sufficient to find a vector field $\mc{V}$ that solves the equation
\begin{equation} \label{eq:VCMatching}
  \mathcal{I}_{ab}(C_{\pm}) = \lie_{[\xi, \mc{V}]}g^{\0}_{ab}
\end{equation}
at the QES. We claim that $\mc{V}$ is precisely the displacement profile of the QES $v$ at $O(\lambda G_N)$ (see eq.\ \eqref{eq:vplus}), i.e.
\begin{equation}
    \mc{V}_{\pm}(0, 0, y) = v_{\pm}(0, 0, y).
\end{equation}
Of course, this is exactly what one should expect, since $\mc{V}$ is precisely the vector field that transforms the metric $\langle h \rangle$ to the Hollands-Wald gauge. For brevity we will only show this for the contour $C_-$; the contour $C_+$ can be analyzed similarly.

Recall that $\mc{I}_{++}(C_-)$ is delta function localized on the past null horizon $\mc{H}^-$. To extract the coefficient of the delta function, we integrate both sides with respect to $x^+$ on the surface $x^- = 0$. This yields
\begin{equation}
    \begin{split}
        \int_0^\infty dx^+ h_{++}(x^+, 0, y) &= \int_0^\infty dx^+ \lie_{[\xi, \mc{V}]} g^\0_{++}(x^+, 0, y) \\
        &= \pi \int_{0}^{\infty} dx^+ \left( x^+ \partial_+^2 \mc{V}_{+}(x^+, 0, y) + 2 \partial_+ \mc{V}_{+}(x^+, 0, y)\right),
    \end{split}
\end{equation}
where, in the second line, we have expanded the Lie derivative.
Finally, integrating the first term in the right hand side by parts and using the condition that $\mc{V}_{+}$ decays at infinity, it is easy to see that we get
\begin{equation}
  \mc{V}_{+}(0,0,y) = \frac{1}{2}\int_0^{\infty} dx^+ \langle h_{++}(x^+,0,y) \rangle.
\end{equation}
This precisely agrees with eq.\ \eqref{eq:vplus}. Proceeding analogously for the contour at $-\infty$, we obtain
\begin{equation}
    (\mc{V}_{\pm} )_{X} = (v_{\pm})_{X},
\end{equation}
which is the desired result.

The only issue here is the extra stress tensor term in eq.\ \eqref{eq:HWmatch}. Naively, this term looks unphysical since it sensitively depends on the vector field $\mc{V}$ on the full homology surface $r$, and the boundary relative entropy should not depend on the values of $\mc{V}$ far away from the QES. Indeed, this term did not appear in the gravitational expression of the boundary relative entropy eq.\ \eqref{eq:SrelMain}. We expect that once the other matter diagrams are processed carefully, their gauge transformation should exactly cancel this unphysical term. In other words, all the other $s$-channel diagrams that contain matter insertions on $r$ should combine with the graviton exchange diagram and produce the full semi-classical symplectic form. We will not carry out this exercise in this paper since our main goal was to simply recover the QES prescription. It would be interesting to see how the full gravitational relative entropy formula eq.\ \eqref{eq:SrelMain} emerges directly from modular Witten diagrams, but we leave this for future work.

\section*{Acknowledgments}
We would like to thank Pawel Caputa, Jan de Boer, Abhijit Gadde, Shiraz Minwalla and Sandip Trivedi for helpful discussions. We acknowledge support from the Department of Atomic Energy, Government of
India, under project identification number RTI 4002, and from the Infosys Endowment for
the study of the Quantum Structure of Spacetime.

\appendix
\section{Coordinates and Conventions}\label{app:A}
\textbf{Index Conventions.} We use lowercase Latin letters $a$, $b$, $c,\ d,\ \cdots$ to denote full spacetime indices. Uppercase Latin letters $I,\ J,\ K,\ L,\ M,\ \cdots$ denote coordinates orthogonal to the extremal surface, and Greek indices $\mu,\ \nu,\ \lambda, \alpha, \beta,\ \cdots$ denote coordinates along the extremal surface. We choose coordinates such that the extremal surface lies at $x^I = 0$.

\noindent\textbf{Rindler Coordinates.} Recall that the standard AdS Rindler wedge metric is given by
\begin{equation}
    g^\0 = -(r^2 - 1)dt^2 + \frac{dr^2}{r^2 - 1} + \frac{r^2}{u^2} (du^2 + d\mathbf{x}^2).
\end{equation}
where $r \geq 1$, $t \in \RR$, and $(u,\mathbf{x})$ parameterize a $(d-1)$ dimensional hyperboloid. Note that the vector field $\xi = 2\pi \partial_t$ is a Killing vector of the above metric. This system only covers one of the two Rindler wedges.

\noindent\textbf{Light cone coordinates.} We define surface adapted light cone coordinates via the relations $x^\pm = \sqrt{r^2 - 1}e^{\pm t}$. Clearly, $x^\pm \geq 0$ is equivalent to the usual Rindler coordinate patch described above; the complementary Rindler wedge can be obtained by continuing $x^\pm$ to negative values. In these coordinates the metric becomes
\begin{align} \label{eq:RindlerLCMet}
    g^\0 =\ &-\frac{1}{4(1+x^+x^-)}\left[(x^-)^2 (dx^+)^2 + (x^+)^2 (dx^-)^2\right] \nonumber\\ &+ \left[\frac{1}{4} + \frac{1}{4(1 + x^+x^-)}\right](dx^+ dx^- + dx^- dx^+) 
    + \frac{1 + x^+x^-}{u^2}(du^2 + d\mathbf{x}^2).
\end{align}
The inverse metric given by
\begin{equation} \label{eq:RindlerLCInvMet}
    (g^\0)^{-1} = (x^+)^2 \partial_{+}^2 + (x^-)^2 \partial_-^2 + (2 + x^+x^-)(\partial_+ \partial_- + \partial_- \partial_+) + \frac{u^2}{1+x^+x^-}(\partial_u^2 + \partial_{\mathbf{x}}^2).
\end{equation}
Likewise, the Killing vector $\xi$ is given by 
\begin{equation}
    \xi = 2\pi(x^+ \partial_+ - x^- \partial_-)  
\end{equation}
Below, we note down some useful Christoffel symbols:
\begin{align}
    \Gamma^{+}_{++} &= \frac{x^+ (x^-)^2}{4(1 + x^+ x^-)} & \Gamma^{-}_{--} &= \frac{x^- (x^+)^2}{4(1 + x^+ x^-)} \label{eq:CS1} \\ 
    \Gamma^{+}_{+-} &= -\frac{x^+ (2 + x^+ x^-)}{4(1 + x^+ x^-)} & \Gamma^{-}_{-+} &= -\frac{x^-(2 + x^+ x^-)}{4(1 + x^+x^-)} \label{eq:CS2} \\
    \Gamma^{+}_{--} &= \frac{(x^+)^3}{4(1 + x^+ x^-)} & \Gamma^{-}_{++} &= \frac{(x^-)^3}{4(1 + x^+ x^-)} \label{eq:CS3} \\
    \Gamma^{+}_{\alpha \alpha} &= -\frac{x^+(1 + x^+ x^-)}{u^2} & \Gamma^{-}_{\alpha \alpha} &= -\frac{x^-(1 + x^+ x^-)}{u^2} \label{eq:CS4} \\
    \Gamma^{\alpha}_{\alpha +} & = \frac{x^+}{2(1 + x^+ x^-)} & \Gamma^{\alpha}_{\alpha -} &= \frac{x^-}{4(1 + x^+ x^-)} \label{eq:CS5}
\end{align}

\section{A brief review of Connes cocycle flow} \label{app:CC}
In this appendix, we will provide a quick review of the Connes cocycle and related constructions in Tomita-Takesaki theory. The interested reader can find more detailed treatments in \cite{takesaki2003theory, bratteli2012operator}.

Let $D(R)$ be the domain of dependence of some spacelike subregion $R$ and let $\mc{M}$ be the associated von Neumann algebra of operators. Let $\mc{H}$ be the Hilbert space on which $\mc{M}$ acts, and let $\Omega \in \mc{H}$ be a cyclic and separating vector for $\mc{M}$.\footnote{Recall that a state $\Psi$ is cyclic for a von Neumann algebra $\mc{M}$ if the set $\{a|\Psi\rangle\, |\, a \in \mc{M}\}$ is dense in Hilbert space. Likewise, it is said to be separating for $\mc{M}$ if $a|\Psi\rangle = 0$ implies that $a = 0$.} For concreteness, we may take $\Omega$ to be the vacuum state which we know to be cyclic and separating by virtue of the Reeh-Schlieder property \cite{Reeh:1961ujh}. The \emph{Tomita operator} $S_\Omega$ is defined as the (anti-linear) map
\begin{equation}
    S_\Omega a|\Omega\rangle = a^\dagger |\Omega\rangle.
\end{equation}
for all $a \in \mc{M}$. Clearly, it follows from definition that $\Omega$ is invertible and densely defined, and can therefore the extended to an (unbounded) operator on $\mc{H}$. Performing a polar decomposition on $S_{\Omega}$ we get
\begin{eqnarray}
    S_\Omega = J_\Omega \Delta_{\Omega}^{1/2},
\end{eqnarray}
where $J_\Omega$ is an anti-unitary operator called \emph{modular conjugation} and $\Delta_\Omega = S^\dagger_\Omega S_\Omega$ is called the \emph{modular operator}. The modular operator for the commutant algebra $\mc{M}'$ is given by
\begin{equation}
    \Delta^\prime_{\Omega} = \Delta_{\Omega}^{-1}.
\end{equation}
It can be shown that $J_\Omega^2 = 1$, and that it sends the algebra $\mc{M}$ to its commutant $\mc{M}'$, i.e. $J_\Omega \mc{A} J_{\Omega} = \mc{A}^\prime$. The fundamental theorem of Tomita-Takesaki theory is that the operator $\Delta_{\Omega}^{is}$ generates an automorphism of the algebra called modular flow. However, this is an outer automorphism in that $\Delta_{\Omega} \notin \mc{M}$. Indeed, the Bisognano-Wichmann theorem \cite{Bisognano:1975ih} states that the generator of modular flow in the vacuum state for the half-space cut in flat space is the boost generator on a \emph{complete} Cauchy surface. This begs the question if there exists an inner automorphism of $\mc{M}$ generated by a local unitary $u \in \mc{M}$. The Connes cocycle is such a unitary.


To define the cocycle we first need to introduce the notion of relative modular flow. Let $\Psi$ and $\Omega$ be two cyclic and separating vectors. For concreteness, we may think of $\Omega$ as the vacuum and $\Psi$ as some excited state. The \emph{relative Tomita operator} $S_{\Psi|\Omega}$ is defined as\footnote{In fact, the assumption that $\Psi$ and $\Omega$ be cyclic and separating is not necessary. For more general definitions see \cite{Ceyhan:2018zfg}.}
\begin{equation}
    S_{\Psi | \Omega} a |\Omega\rangle = a^\dagger|\Psi\rangle.
\end{equation}
Clearly, $S_{\Psi|\Omega}$ is also invertible, densely defined, and therefore closable.  Moreover, one can recover the Tomita operator from the relative Tomita operator as $S_{\Omega | \Omega} = S_{\Omega}$. Performing a polar decomposition we can rewrite $S_{\Psi|\Omega}$ as
\begin{eqnarray}
    S_{\Psi|\Omega} = J_{\Psi|\Omega} \Delta^{1/2}_{\Psi|\Omega}
\end{eqnarray}
where $J_{\Psi | \Omega}$ is anti-unitary and $\Delta_{\Psi|\Omega} := S_{\Psi | \Omega}^\dagger S_{\Psi|\Omega}$ is positive. The operator $\Delta_{\Psi | \Omega}$ is called the \emph{relative modular operator}, and the corresponding unitary flow generated by $\Delta_{\Psi | \Omega}^{is}$ is called relative modular flow. The relative modular operator for the commutant $\mc{M}^\prime$ is
\begin{equation} \label{eq:CCdef}
    \Delta^\prime_{\Psi | \Omega} = \Delta^{-1}_{\Omega|\Psi}.
\end{equation}
The Connes cocycle $u_{\Psi | \Omega}(s)$ is then defined
\begin{eqnarray}
    u_{\Psi | \Omega}(s) = \Delta_{\Psi | \Omega}^{is} \Delta_{\Omega}^{-is}
\end{eqnarray}
The cocycle on the complementary algebra is defined analogously. One can show that $u_{\Psi | \Omega}(s)$ generates an inner automorphism of $\mc{M}$ \cite{connes1978homogeneity, Araki:1973hh} (see also \cite{Lashkari:2019ixo}). Moreover, it is easy to check from definition that the cocycle intertwines modular flows for $\Psi$ and $\Omega$:
\begin{equation}
    \Delta_{\Psi}^{is} = u_{\Psi | \Omega}(s) \Delta_{\Omega}^{is} u^{\dagger}_{\Psi | \Omega}(s) \label{eq:CCinter}.
\end{equation}
Let $\mathsf{a} \in \mc{M}$ and $\mathsf{b'} \in \mc{M}'$. Then it can be shown that the cocycle acts as one-sided boosts at the level of correlation functions evaluated in the excited state $\Psi$:
\begin{align}
    \langle \Psi | u^{\dagger}_{\Psi | \Omega}(s) \mathsf{a} u_{\Psi | \Omega}(s) | \Psi \rangle &= \langle \Psi | \Delta_{\Omega}^{is} \mathsf{a} \Delta_{\Omega}^{-is} | \Psi \rangle \label{eq:CCM} \\
    \langle \Psi | u^{\dagger}_{\Psi | \Omega}(s) \mathsf{b}' u_{\Psi | \Omega}(s) | \Psi \rangle &= \langle \Psi | \mathsf{b}' | \Psi \rangle \label{eq:CCMcomm}.
\end{align}
All of the above properties are straightforward to verify if the algebra $\mc{M}$ is type I. Let $\rho_{\Psi}$ and $\rho'_{\Psi}$ (similarly $\rho_{\Omega}$ and $\rho_{\Omega}'$) be the reduced density matrices on $\mc{M}$ and $\mc{M}'$ respectively. Then one has
\begin{align}
    \Delta_{\Omega} &= \rho_{\Omega} (\rho'_{\Omega})^{-1} \\
    \Delta_{\Psi | \Omega} &= \rho_{\Psi} (\rho_{\Omega}')^{-1} \\
    u_{\Psi | \Omega}(s) &= \rho_{\Psi}^{is} \rho_{\Omega}^{-is},
\end{align}
and eqs.\ \eqref{eq:CCinter}, \eqref{eq:CCM} and \eqref{eq:CCMcomm} follow directly.



\section{Connes cocycle flow in AdS$_3$} \label{app:CCBTZ}
\begin{figure}[t]
    \centering
    \begin{subfigure}{0.45\textwidth}
        \centering
        \includegraphics[width=\textwidth]{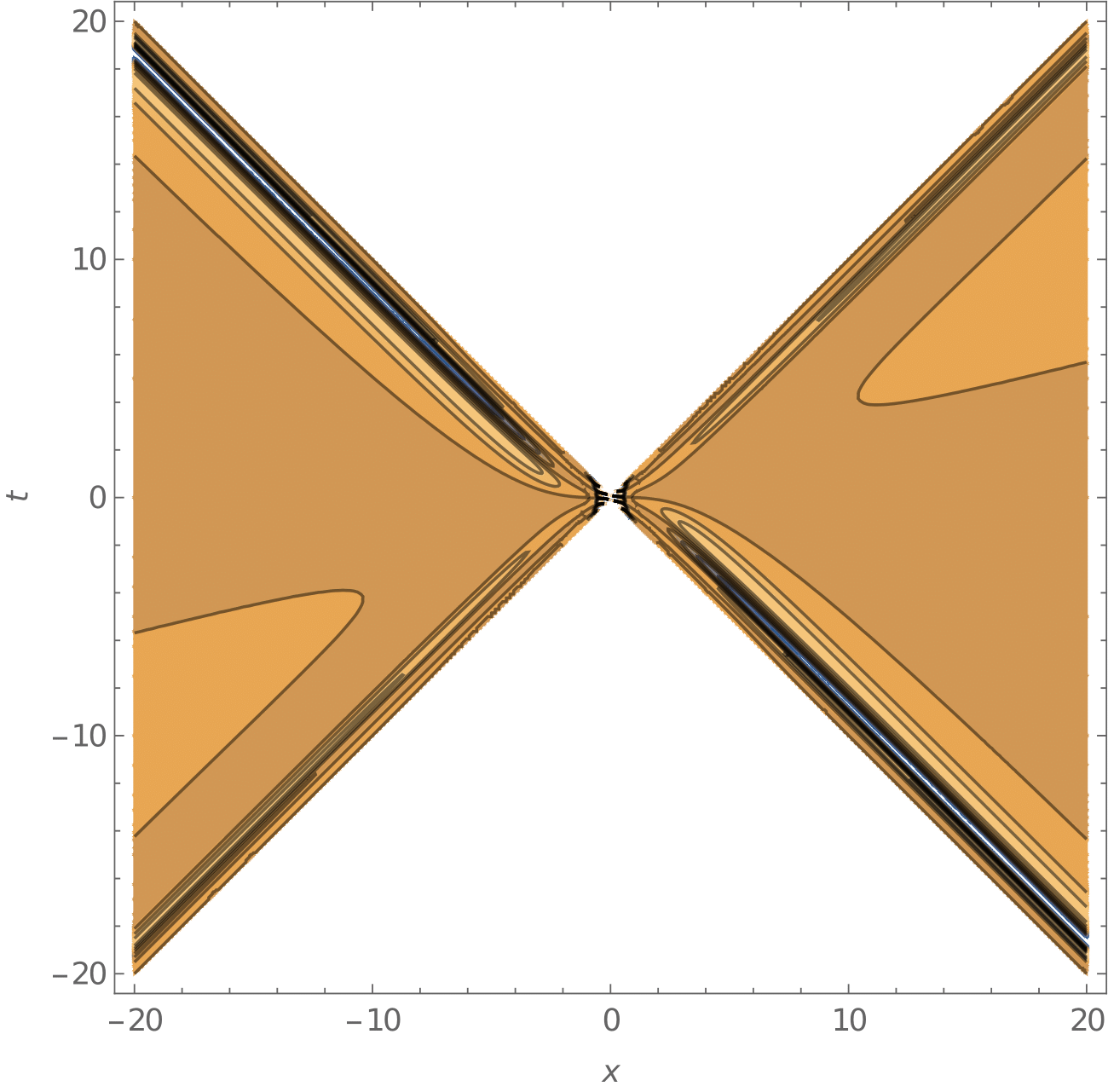}
        \caption{$s = 0$}
    \end{subfigure}
    \hfill
    \begin{subfigure}{0.45\textwidth}
        \centering
        \includegraphics[width=\textwidth]{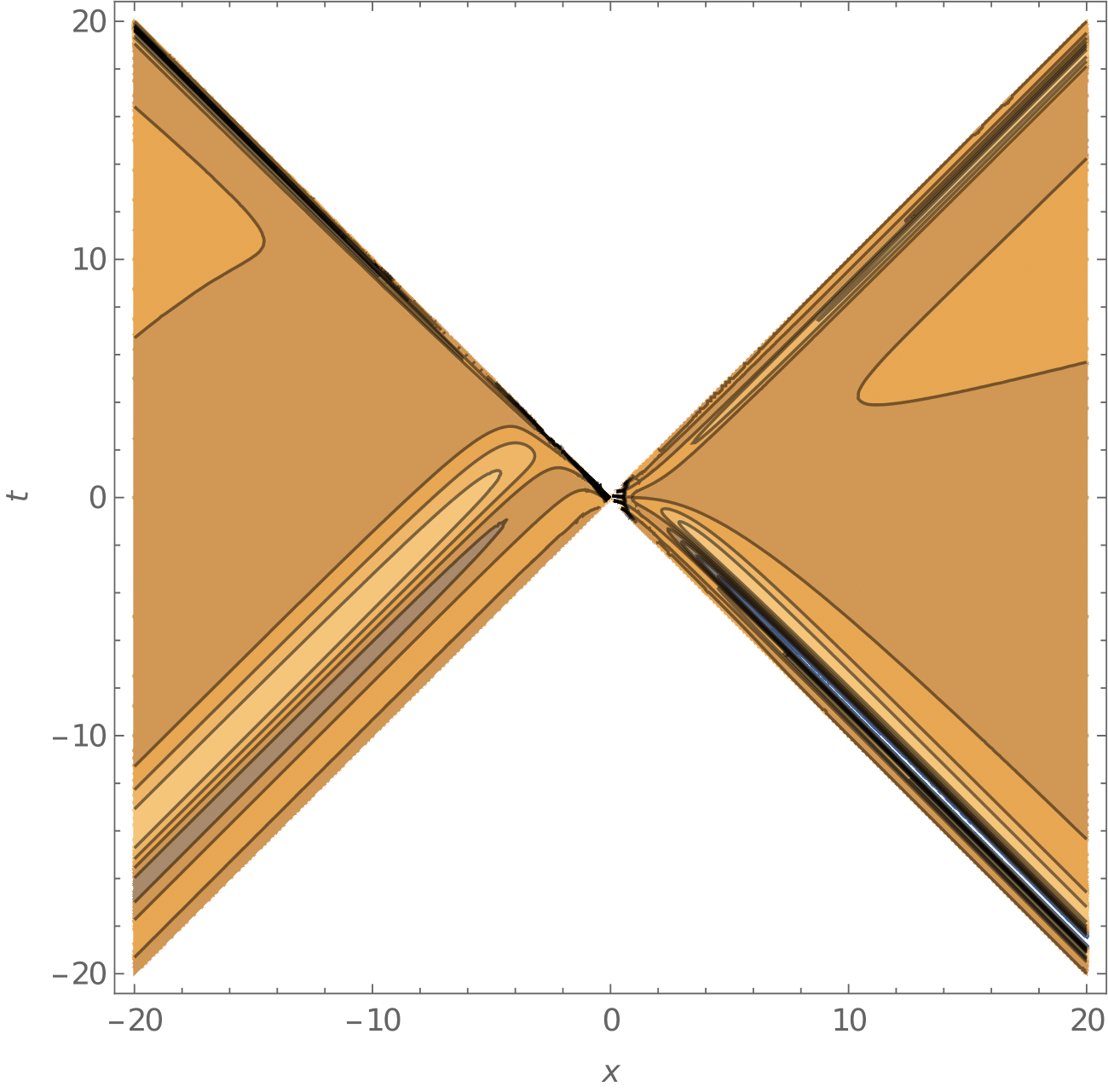}
        \caption{$s = 1.5$}
    \end{subfigure}
    \hfill
    \vspace{11pt}
    \begin{subfigure}{0.45\textwidth}
        \centering
        \includegraphics[width=\textwidth]{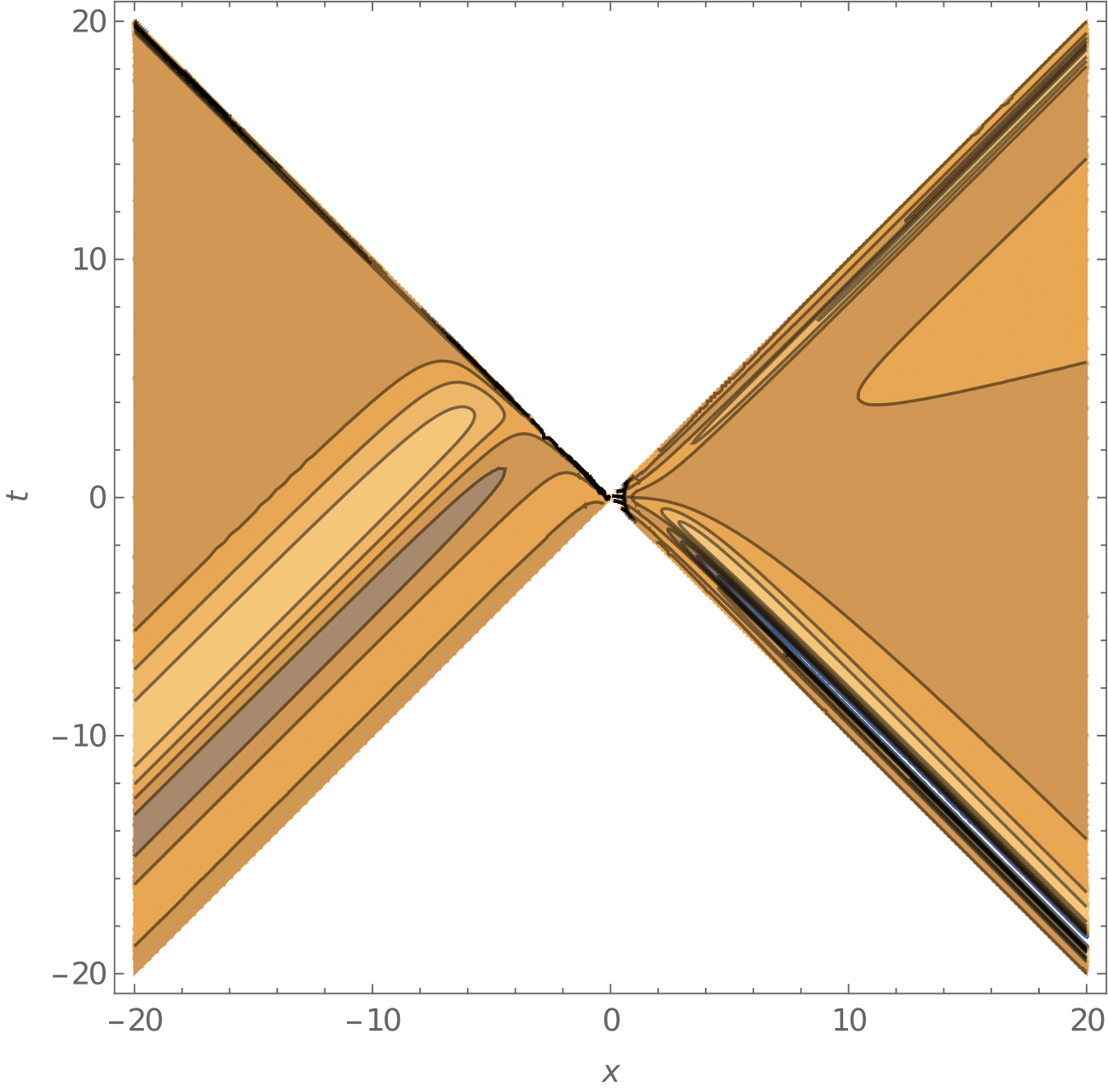}
        \caption{$s = 2$}
    \end{subfigure}
    \hfill
    \begin{subfigure}{0.45\textwidth}
        \centering
        \includegraphics[width=\textwidth]{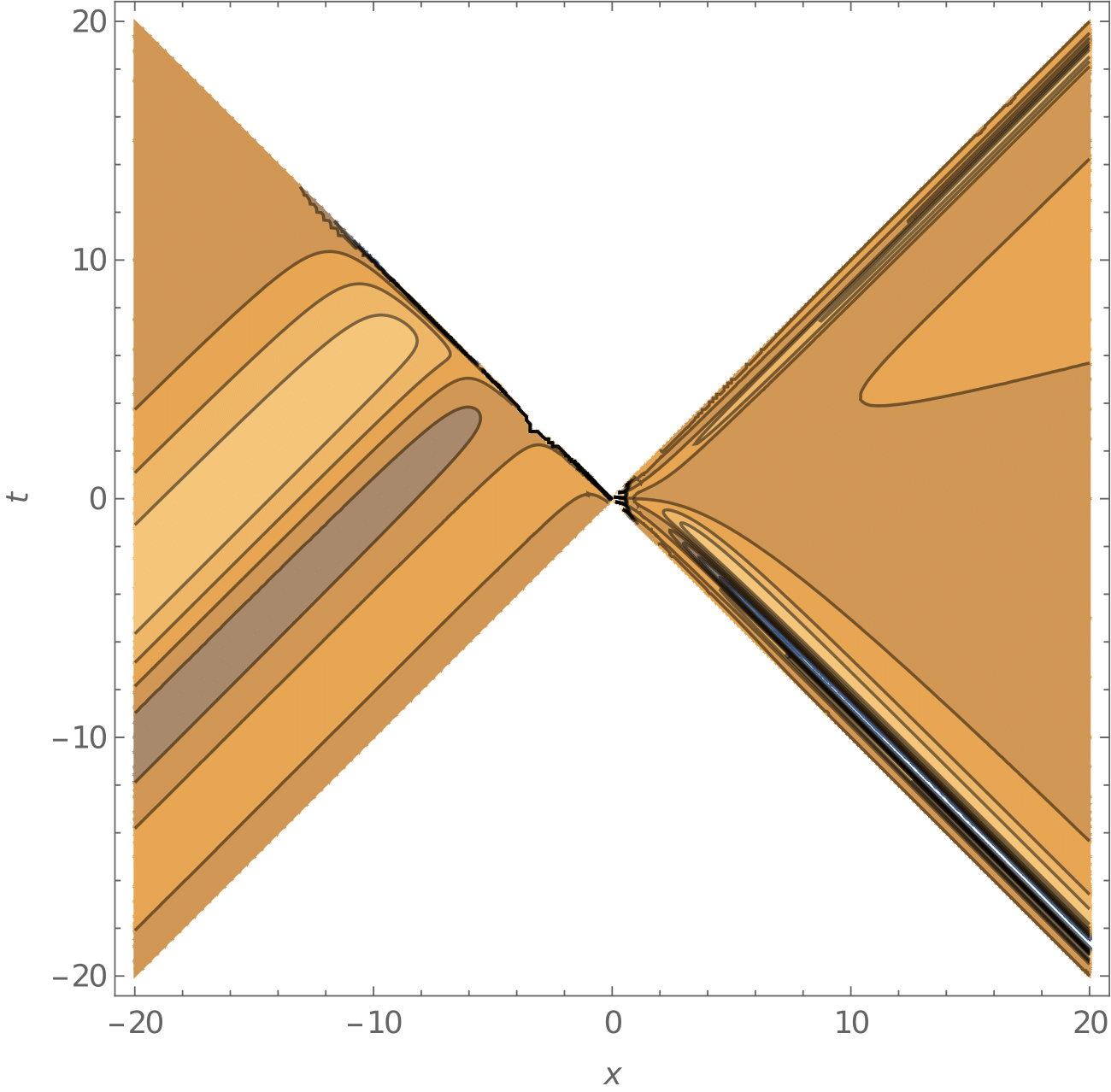}
        \caption{$s = 2.5$}
    \end{subfigure}
    \caption{Plotes of $T_{++}$ restricted to the left and right Rindler wedges with varying values of the one-sided boost parameter $s$.}
    \label{fig:CC3}
\end{figure}

In this Appendix, we will illustrate the effect of CC flow on $T_{++}$ in AdS$_3$. The metric of global AdS$_3$ can be written as
\begin{equation}
    ds^2 = \frac{4}{1 + t^2 - x^2}(- dt^2 + dx^2) + \left(\frac{1 - t^2 + x^2}{1 + t^2 - x^2}\right) d\phi^2
\end{equation}
where $t, x,\phi \in \RR$.\footnote{Compactifying the $\phi$ coordinate leads to the BTZ black hole solution.} The AdS Rindler wedges are defined by the conditions $|x| \geq |t|$, with the left wedge corresponding to the region $x < 0$ and the right wedge corresponding to the region $x > 0$. Note that the QES lives at $x = t = 0$. This metric has a \lq\lq boost\rq\rq\, symmetry generated by the transformations
\begin{align}
    t \rightarrow t(s) &= \sinh(s) t + \cosh(s) x \\
    x \rightarrow x(s) &= \cosh(s) t + \sinh(s) x.
\end{align}
Let us assume that the bulk matter state is the Hartle-Hawking wave function. The bulk one-point function of a bulk scalar field $\psi$ due to the insertion of an operator in the Euclidean boundary is given by \cite{Maldacena:2001kr}
\begin{equation}
    \langle \psi(t, x, \phi) \rangle \sim \left[\frac{1 + t^2 - x^2}{(1 - t^2 + x^2)(\cosh(2\pi(\phi - \phi_0)) - 1) + (t + x - e^{-i\theta_0})(t - x + e^{i\theta_0})}\right]^{2\Delta}
\end{equation}
where $(\theta_0, \phi_0)$ is the coordinate location of the boundary operator insertion, $\Delta$ is its conformal dimension, and we have set $\beta = 1$. For an operator insertion at $\theta_0=-\pi/2$ and $\phi_0 = 0$, the one-point function on the $\phi = 0$ slice becomes
\begin{equation}
    \langle \psi(t, x, \phi=0) \rangle \sim \left[ \frac{1 + t^2 - x^2}{(t + x + i)(t - x + i)} \right]^{2\Delta}.
\end{equation}
We can then compute the associated stress tensor $T_{ab}$ and perform a one-sided boost on the left wedge. Some plots showing the profile of the $++$ component of the stress tensor evaluated in the $\phi = 0$ cross section are given in figure \ref{fig:CC3}. The value of the boost parameter $s$ is increased from $s = 0$ to $s = 2.5$.
Figure \ref{fig:CC3} shows a clear trend --- the boost dilutes out the fluctuations of the stress tensor in the left wedge with the right wedge is held fixed. Consequently, in the strict $s \to \infty$ limit, the quantum state as perceived by any local observer in the left wedge will be indistinguishable from the vacuum state.

\section{Some details on the Hollands-Wald gauge} \label{app:HW}
The first Hollands-Wald gauge condition states that the extremal surface does not move when the boundary double trace operator is switched on. So the deformed QES satisfies the variational equation
\begin{equation}
    \dl_{X} \left[ \frac{1}{4G_N} \dl_g \mc{A}^g + (\dl_g + \dl_\psi) S^{g,\psi}_\text{bulk} \right]_{X^\0} = 0.
\end{equation}
This simply yields the extremality equation \ref{eq:QEE_Rindler} with the displacement profile $v_I$ set to zero. Let $\gamma_{ab}$ be the metric in the Hollands-Wald gauge. The the first gauge fixing condition is
\begin{equation} \label{eq:HWGaugeFixing1}
    \widehat{\nabla}^2 w^{(\gamma)}_+ = 0, \qquad w_{+}^{(\gamma)} = \frac{1}{2}\int_{0}^{\infty} dx^+ \gamma_{++}(x^+, 0, y)
\end{equation}
Now, let $h_{ab}$ be the metric perturbation in an arbitrary gauge so that $\gamma_{ab} = h_{ab} + 2 \nabla_{(a} \mc{V}_{b)}$. The gauge transformation $\mc{V}$ is often referred to as the Hollands-Wald vector field. It is then easy to see that
\begin{equation}
    w_+^{(\gamma)} = w_+^{(h)} - (\mc{V}_+)_{X^\0}
\end{equation}
where $w_+^{(h)}$ has been defined in an analogous fashion. Plugging this into eq. \ref{eq:HWGaugeFixing1}, we arrive at 
\begin{equation}
    \widehat{\nabla}^2 w_+^{(h)}(y) = \widehat{\nabla}^2 \mc{V}_+(y).
\end{equation}
In the absence of a stress tensor insertion at the boundary, the (Lorentzian) metric perturbation $h_{ab}$ asymptotically approaches 0. Setting the boundary condition $\mc{V}_+ \rightarrow 0$ at spatial infinity lets us conclude that
\begin{equation}
    \mc{V}_+(y) = w_+^{(h)}(y) = \frac{1}{2} \int_{0}^\infty dx^+ h_{++}(x^+, 0, y).
\end{equation}
As a corrolary notice that the Hollands-Wald vector field is precisely equal to the displacement profile of the QES. This makes intuitive sense since the diffeomorphism generated by the Hollands-Wald vector field simply moves the undeformed QES to the deformed QES at linearized order.

For the sake of completeness, let us also write down the second Hollands-Wald condition \ref{itm:QHW2} in terms of the vector field $\mc{V}$. The second Hollands-Wald condition states that $\lie_\xi(g^\0 + \gamma)|_{X^\0} = 0$. Since $\xi$ is already a Killing vector of the undeformed metric, this condition can be rewritten as
\begin{equation}
  (\lie_{[\xi, \mc{V}]} g^{\0})_{X^\0} = -(\lie_{\xi} h)_{X^\0}.
\end{equation}
Explicitly working out both sides, we get the following equations for $V$:
\begin{align} \label{eq:HW2V}
  (\partial_{\pm}\mc{V}^\mp)_{X^\0} &= -(h_{\pm\pm})_{X^\0} \\
  (\partial_{\pm} \mc{V}_{\alpha} + \partial_{\alpha} \mc{V}_{\pm})_{X^\0} &= -(h_{\pm\alpha})_{X^\0}
\end{align}
So we see that the first Hollands-Wald condition specifies the value of $\mc{V}$ at the quantum extremal surface, and the second condition specifies its first derivatives.  In this sense the Hollands-Wald gauge fixes $\mc{V}$ only in an infinitesimal neighborhood of the QES.

\section{Feynman rules on the time fold} \label{app:Feyn-rules}
\begin{figure}[t]
    \centering
    \includegraphics[height=3.5cm]{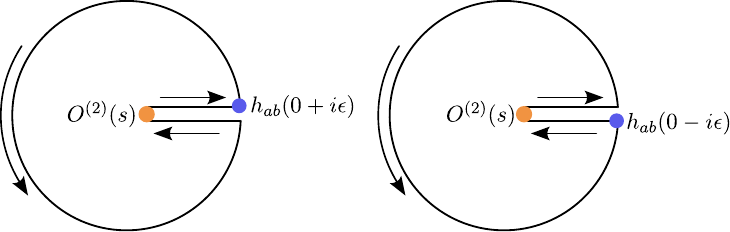}
    \caption{Contour ordered operator insertions appearing in the two terms of the commutator $\langle[h_{ab}(0), O^\2(s)]\rangle$. Here we have only shown the time contour and suppressed the spatial directions.}
    \label{fig:commutator}
\end{figure}

In this appendix, we will illustrate the Feynman rules introduced in section \ref{sec:invHKLL}. Suppose we want to compute the causal bulk-to-boundary propagator given by$\langle [ h_{ab}(0), O^\2(s)] \rangle$ (the spatial coordinates have been suppressed for notational simplicity).  As shown in figure \ref{fig:commutator}, the ordering of operators inside the commutator is enforced by the contour ordering coming from the path integral. We can perturbatively compute the above causal propagator by pulling down interaction vertices order by order everywhere along the contour, including the Euclidean circle. However, it is easy to check that vertices inserted along the Euclidean section do not contribute since all such diagrams cancel between the two terms of the commutator. So it suffices to only consider insertions on the real time fold.

At first order, we only need to pull down a single interaction vertex of the form $\int T^{ab} h_{ab}$ at time $0 < t < s$. The vertex can be inserted on either the top or the bottom branch of the fold. This leads to $2 \times 2 = 4$ terms which combine to give a product of commutators of the form $\int dt\, \langle [T_{ab}(t), O^\2(s)] \rangle \langle [h_{ab}(0), T_{ab}(t) ] \rangle$. This shows us the that the perturbative expansion of the causal bulk-to-boundary propagator can be expressed as a composition of causal propagators, consistent with eq.\ \eqref{eq:KCdef}. This also shows that the interaction vertex need only be integrated over the past causal wedge of $s$ since the commutators vanish outside this wedge. 

If however, we wish to compute Wightman functions and not commutators, we cannot ignore vertex insertions in the Euclidean section of the geometry. All Euclidean insertions will lead to contour ordered Wightman propagators, whereas Lorentzian insertions lead to causal propagators since time flows in opposite directions in the two branches of the fold. In fact, as we saw in the main text, such Euclidean insertions were crucial for us to recover the QES prescription.

\section{Analytic continuation of $\mc{T}_3$ and $\mc{T}_4$} \label{app:T3T4}
In this appendix we will process the diagrams $\mc{T}_3$ and $\mc{T}_4$ of section \ref{sec:grsk-derivation}. Let us first begin with $\mc{T}_3$ which is given by
\begin{equation}
    \mc{T}_3 = \int_{2\pi}^{2\pi\alpha}d\tau_x \int_0^{2\pi} d\tau_y\, K^{ab}_\alpha(\tau_x | \tau) G^\alpha_{abcd}(\tau_x | \tau_y) K^{cd}_{\alpha}(\tau_y | \theta).
\end{equation}
In this diagram, the $\tau_y$ branch cut lives in the regular region whereas $\theta$ lives in the excess region. Consequently, we may analytically continue $\theta$ anywhere in the strip $\mc{S}$ before taking the $\alpha \to 1$ limit without worrying about crossing branch cuts. Next, we wish to continue $\tau_x$. Note that there is no branch cut corresponding to $\theta$ in the complex $\tau_x$ plane --- the two branch cuts come from $\tau_y$ and the boundary insertion at $\tau$. Moreover, both branch cuts live in the regular region, and therefore in the $\alpha \to 1$ limit, the $\tau_x$ integral vanishes. Consequently $\mc{T}_3$ vanishes.

The case of $\mc{T}_4$ is slightly more subtle since both interaction vertices are inserted in the excess region:
\begin{equation}
    \mc{T}_4 = \int_{2\pi}^{2\pi\alpha}d\tau_x \int_{2\pi}^{2\pi\alpha} d\tau_y\, K^{ab}_\alpha(\tau_x | \tau) G^\alpha_{abcd}(\tau_x | \tau_y) K^{cd}_{\alpha}(\tau_y | \theta).
\end{equation}
To process this diagram, we first interchange the $\tau_x$ and $\tau_y$ integrals, and then split up the integrals as follows:
\begin{equation} \label{eq:F3}
    \left(\int_{2\pi}^{\theta} d\tau_y + \int_{\theta}^{2\pi \alpha} d\tau_y \right) \left( \int_{2\pi}^{\tau_y} d\tau_x + \int_{\tau_y}^{2\pi \alpha} d\tau_x\right) K^{ab}_\alpha(\tau_x | \tau) G^\alpha_{abcd}(\tau_x | \tau_y) K^{cd}_{\alpha}(\tau_y | \theta).
\end{equation}
As before, since $\tau_y$ lies in the excess region, we continue $\theta$ depending on whether $\theta > \tau_y$ or $\theta < \tau_y$. If $\theta > \tau_y$, we send $\theta \to 2\pi \alpha + is$, otherwise we send $\theta \to 2\pi + is$. We then continue the $\tau_y$ contour according to the prescription shown in figure \ref{fig:bulk-cont}. 

Next, we wish to continue the $\tau_x$ contour. The idea here is the same as before --- we continue the endpoints of the $\tau_x$ integrals to Lorentzian values in such a way that no branch cuts are crossed. To see how this goes, consider the terms
\begin{equation} \label{eq:F4}
    \left( \int_{\gamma_1} d\tau_y \int_{2\pi}^{\tau_y} d\tau_x + \int_{\gamma_1} d\tau_y \int_{\tau_y}^{2\pi\alpha} d\tau_x \right) K^{ab}_\alpha(\tau_x | \tau) G^\alpha_{abcd}(\tau_x | \tau_y) K^{cd}_{\alpha}(\tau_y | \theta).
\end{equation}
The contour $\gamma_1$ (and $\gamma_2$) contains a small Euclidean segment that vanishes in the $\alpha \to 1$ limit; in the following we will ignore this segment and only focus on the Lorentzian part running from $2\pi \alpha$ to $2\pi \alpha + is$ . In the complex $\tau_x$ plane, the $\tau_y$ branch cut lies on the upper boundary of $\mc{S}$ in both of the terms shown above. Now, in the first term, if we imagine that $\tau_y$ is Euclidean, then $\tau_x < \tau_y$, and therefore the endpoint point of the $\tau_x$ contour must be continued in such a way that it lies below the $\tau_y$ branch cut. Likewise, in the second term, the endpoint must lie above the branch cut. Therefore, as we continue $\tau_y$ to Lorentzian values, we simultaneously arrive at the following analytic continuation of $\tau_x$:
\begin{equation}
    \left(\int_{2\pi \alpha}^{2\pi \alpha + is} d\tau_y \int_{2\pi \alpha - \epsilon}^{2\pi \alpha - \epsilon + is} d\tau_x + \int_{2\pi \alpha}^{2\pi \alpha + is} d\tau_y \int_{2\pi \alpha + \epsilon + is}^{2\pi \alpha + \epsilon} d\tau_x\right)K^{ab}_\alpha(\tau_x | \tau) G^\alpha_{abcd}(\tau_x | \tau_y) K^{cd}_{\alpha}(\tau_y | \theta),
\end{equation}
where we have dropped the small vertical Euclidean segment in both the $\tau_x$ and $\tau_y$ integrals. Consequently, in the $\alpha \to 1$ limit, the bulk-to-bulk propagator $G^{\alpha}_{abcd}$ turns into a commutator. Performing similar manipulations on the other two terms of eq.\ \eqref{eq:F3}, we observe that the appropriate analytic continuation of $\mc{T}_4$ is given by
\begin{equation}
    \int_{0}^{s} ds_x \int_{0}^{s} ds_y\, K^{ab}(is_x | \tau) G_{C;abcd}(s_x | s_y) K_C^{cd}(s_y | s),
\end{equation}
which is exactly the SK diagram with both interaction vertices inserted in the Lorentzian section.

\bibliography{refs}
\bibliographystyle{JHEP}

\end{document}